\documentclass[twocolumn, fleqn]{doc_class}

\usepackage[super, comma, sort&compress]{natbib}

\usepackage{multibib}
\newcites{m}{References}
\bibliographystylem{bib_style}
\newcites{sup}{References (continued)}
\bibliographystylesup{bib_style}

\usepackage{newtxtext, newtxmath}
\usepackage[T1]{fontenc}
\usepackage[utf8]{inputenc}

\usepackage[dvipsnames, hyperref]{xcolor}
\usepackage{ae, aecompl}
\usepackage{graphicx}
\usepackage{xspace}
\usepackage{soul}
\usepackage{booktabs}
\usepackage{adjustbox}
\usepackage{comment}

\usepackage{amsmath}
\usepackage{commath}
\usepackage{siunitx}
\usepackage{nth}

\usepackage[capitalise, noabbrev]{cleveref}
\usepackage{csvsimple}
\usepackage{multirow}
\usepackage{array}
\newcolumntype{P}[1]{>{\centering\arraybackslash}p{#1}}

\crefname{figure}{Fig.}{Figs.}
\crefname{equation}{equation}{equations}

\usepackage[leftmargin=0pt,rightmargin=0pt,skipabove=2pt,skipbelow=2pt,innerleftmargin=5pt,innerrightmargin=5pt,innertopmargin=5pt,innerbottommargin=5pt,linewidth=0pt,innerlinewidth=0pt,middlelinewidth=0pt,outerlinewidth=0pt,roundcorner=0pt]{mdframed}

\newcommand{\orcidsymb}[2]{#1\href{http://orcid.org/#2}{\adjustbox{trim={-.15\width} {0\height} {-.15\width} {0\height},clip}{\includegraphics[height=10pt]{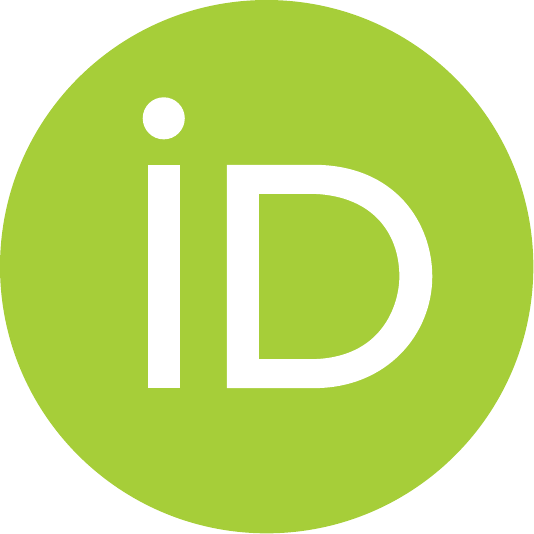}}}}

\hypersetup{
    unicode=true,
    final=true,
    plainpages=false,
    pdfstartview=FitV,
    pdftoolbar=false,
    pdfmenubar=true,
    bookmarksopen=true,
    bookmarksnumbered=true,
    breaklinks=true,
    linktocpage,
    colorlinks=true,
    linkcolor=Blue,
    urlcolor=Blue,
    citecolor=Blue,
    anchorcolor=green
}

\AtBeginDocument{
    \hypersetup{
        pdftitle={Photometric detection at $7.7$ microns of a galaxy beyond redshift $14$ with JWST/MIRI},
        pdfauthor={J. M. Helton, G. H. Rieke, et al.},
        pdfsubject={},
        pdfkeywords={{galaxy evolution} -- {galaxy formation} -- {high-redshift galaxies}}
    }
}

\usepackage{paralist}
\makeatletter
    \def\@biblabel#1{\@ifnotempty{#1}{#1}}
    \def\NAT@anchor#1#2{
        \hfilneg\hyper@natanchorstart{#1\@extra@b@citeb}
        #2.
        \hyper@natanchorend
    }
\makeatother

\renewenvironment{thebibliography}[1]{%
    \newpage
    \section*{\normalsize \refname}
    \setlength{\parindent}{0pt}
    \inparaenum
}{\endinparaenum}

\DeclareRobustCommand{\VAN}[3]{#2}
\let\VANthebibliography\thebibliography
\def\thebibliography{\DeclareRobustCommand{\VAN}[3]{##3}\VANthebibliography}

\title[Photometric detection at \boldmath$7.7\ \mu\mathrm{m}$ of a galaxy beyond redshift \boldmath$14$ with JWST/MIRI]{Photometric detection at \boldmath$7.7\ \mu\mathrm{m}$ of a galaxy beyond redshift \boldmath$14$ with JWST/MIRI}

\author[]{
        {\orcidsymb{Jakob\ M.\ Helton}{0000-0003-4337-6211}$^{\hyperlink{inst:Steward}{1}}$\thanks{E-mail: \href{mailto:jakobhelton@arizona.edu}{jakobhelton@arizona.edu}}, \orcidsymb{George\ H.\ Rieke}{0000-0003-2303-6519}$^{\hyperlink{inst:Steward}{1}}$, \orcidsymb{Stacey\ Alberts}{0000-0002-8909-8782}$^{\hyperlink{inst:Steward}{1}}$, \orcidsymb{Zihao\ Wu}{0000-0002-8876-5248}$^{\hyperlink{inst:CfA}{2}}$, 
    }
    \newauthor{
        \orcidsymb{Daniel\ J.\ Eisenstein}{0000-0002-2929-3121}$^{\hyperlink{inst:CfA}{2}}$, \orcidsymb{Kevin\ N.\ Hainline}{0000-0003-4565-8239}$^{\hyperlink{inst:Steward}{1}}$, \orcidsymb{Stefano\ Carniani}{0000-0002-6719-380X}$^{\hyperlink{inst:SNS}{3}}$, \orcidsymb{Zhiyuan\ Ji}{0000-0001-7673-2257}$^{\hyperlink{inst:Steward}{1}}$, 
    }
    \newauthor{
        \orcidsymb{William\ M.\ Baker}{0000-0003-0215-1104}$^{\hyperlink{inst:Kavli}{4}, \hyperlink{inst:Cav}{5}}$, \orcidsymb{Rachana\ Bhatawdekar}{0000-0003-0883-2226}$^{\hyperlink{inst:ESAC}{6}}$, \orcidsymb{Andrew\ J.\ Bunker}{0000-0002-8651-9879}$^{\hyperlink{inst:Oxford}{7}}$, \orcidsymb{Phillip\ A.\ Cargile}{0000-0002-1617-8917}$^{\hyperlink{inst:CfA}{2}}$, 
    }
    \newauthor{
        \orcidsymb{St\'{e}phane\ Charlot}{0000-0003-3458-2275}$^{\hyperlink{inst:IAP}{8}}$, \orcidsymb{Jacopo\ Chevallard}{0000-0002-7636-0534}$^{\hyperlink{inst:Oxford}{7}}$, \orcidsymb{Francesco\ D'Eugenio}{0000-0003-2388-8172}$^{\hyperlink{inst:Kavli}{4}, \hyperlink{inst:Cav}{5}}$, \orcidsymb{Eiichi\ Egami}{0000-0003-1344-9475}$^{\hyperlink{inst:Steward}{1}}$, 
    }
    \newauthor{
        \orcidsymb{Benjamin\ D.\ Johnson}{0000-0002-9280-7594}$^{\hyperlink{inst:CfA}{2}}$, \orcidsymb{Gareth\ C.\ Jones}{0000-0002-0267-9024}$^{\hyperlink{inst:Oxford}{7}}$, \orcidsymb{Jianwei\ Lyu}{0000-0002-6221-1829}$^{\hyperlink{inst:Steward}{1}}$, \orcidsymb{Roberto\ Maiolino}{0000-0002-4985-3819}$^{\hyperlink{inst:Kavli}{4}, \hyperlink{inst:Cav}{5}, \hyperlink{inst:UCL}{9}}$, 
    }
    \newauthor{
        \orcidsymb{Pablo\ G.\ P\'{e}rez-Gonz\'{a}lez}{0000-0003-4528-5639}$^{\hyperlink{inst:CAB}{10}}$, \orcidsymb{Marcia\ J.\ Rieke}{0000-0002-7893-6170}$^{\hyperlink{inst:Steward}{1}}$, \orcidsymb{Brant Robertson}{0000-0002-4271-0364}$^{\hyperlink{inst:UCSC}{11}}$, \orcidsymb{Aayush\ Saxena}{0000-0001-5333-9970}$^{\hyperlink{inst:Oxford}{7}, \hyperlink{inst:UCL}{9}}$, 
    }
    \newauthor{
        \orcidsymb{Jan\ Scholtz}{0000-0001-6010-6809}$^{\hyperlink{inst:Kavli}{4}, \hyperlink{inst:Cav}{5}}$, \orcidsymb{Irene\ Shivaei}{0000-0003-4702-7561}$^{\hyperlink{inst:CAB}{10}}$, \orcidsymb{Fengwu\ Sun}{0000-0002-4622-6617}$^{\hyperlink{inst:Steward}{1}, \hyperlink{inst:CfA}{2}}$, \orcidsymb{Sandro\ Tacchella}{0000-0002-8224-4505}$^{\hyperlink{inst:Kavli}{4}, \hyperlink{inst:Cav}{5}}$, \orcidsymb{Lily\ Whitler}{0000-0003-1432-7744}$^{\hyperlink{inst:Steward}{1}}$, 
    }
    \newauthor{
        \orcidsymb{Christina\ C.\ Williams}{0000-0003-2919-7495}$^{\hyperlink{inst:NOIRLab}{12}}$, \orcidsymb{Christopher\ N.\ A.\ Willmer}{0000-0001-9262-9997}$^{\hyperlink{inst:Steward}{1}}$, \orcidsymb{Chris Willott}{0000-0002-4201-7367}$^{\hyperlink{inst:NRC}{13}}$, \orcidsymb{Joris\ Witstok}{0000-0002-7595-121X}$^{\hyperlink{inst:Kavli}{4}, \hyperlink{inst:Cav}{5}}$, 
    }
    \newauthor{
        and \orcidsymb{Yongda\ Zhu}{0000-0003-3307-7525}$^{\hyperlink{inst:Steward}{1}}$
    }
    \\
    \\
    {\normalsize Affiliations are listed at the end of the manuscript.}
}

\pubyear{2025}

\begin{document}
\maketitle

% Abstract of the paper
\begin{abstract}
    \begin{mdframed}[backgroundcolor=black!5]
        The James Webb Space Telescope (JWST) has spectroscopically confirmed numerous galaxies at $z > 10$. While weak rest-frame ultraviolet emission lines have only been seen in a handful of sources, the stronger rest-frame optical emission lines are highly diagnostic and accessible at mid-infrared wavelengths with the Mid-Infrared Instrument (MIRI) of JWST. We report the photometric detection of the distant spectroscopically confirmed galaxy JADES-GS-z14-0 at $z = 14.32^{+0.08}_{-0.20}$ with MIRI at $7.7\ \mu\mathrm{m}$. The most plausible solution for the stellar population properties is that this galaxy contains half a billion solar masses in stars with a strong burst of star formation in the most recent few million years. For this model, at least one-third of the flux at $7.7\ \mu\mathrm{m}$ comes from the rest-frame optical emission lines $\mathrm{H}\beta$ and/or $\mathrm{[OIII]}\lambda\lambda4959,5007$. The inferred properties of JADES-GS-z14-0 suggest rapid mass assembly and metal enrichment during the earliest phases of galaxy formation. This work demonstrates the unique power of mid-infrared observations in understanding galaxies at the redshift frontier.
    \end{mdframed}
\end{abstract}


%%%%%%%%%%%%%%%%%%%%%%%%%%%%%%%%%%%%%%%%%%%%%%%%%%

%%%%%%%%%%%%%%%%% BODY OF PAPER %%%%%%%%%%%%%%%%%%

With the launch of the James Webb Space Telescope (JWST), extragalactic astronomy fundamentally changed. The Near Infrared Camera (NIRCam) shifted the photometric redshift frontier from $z \approx 10$ to $z \approx 14-16$ \citem{Castellano:2022, Finkelstein:2023, Harikane:2023, Robertson:2023, Robertson:2024, Hainline:2024}, while the Near Infrared Spectrograph (NIRSpec) pushed the spectroscopic redshift frontier from $z \approx 8$ to $z \approx 12-14$ \citem{Curtis-Lake:2023, Fujimoto:2023, Wang:2023, Castellano:2024}. Crucially, JWST discovered an early period of galaxy formation that was more vigorous than expected, with a sizable population of luminous galaxies and supermassive black holes less than a billion years after the Big Bang.

A companion paper reports the spectroscopic confirmation of JADES-GS-z14-0 at redshift $z = 14.32^{+0.08}_{-0.20}$, which makes it the most distant galaxy with a spectroscopically confirmed redshift \citem{Carniani:2024}. This galaxy is remarkably luminous, with a rest-frame ultraviolet (UV) absolute magnitude $M_{\mathrm{UV}} \approx -20.81 \pm 0.16$, which may require a re-assessment of ideas about early galaxy formation, suggesting a slow decline in the number density of galaxies at $z > 12$, with increasing efficiency of galaxy formation in halos at higher redshifts \citem{Robertson:2024}. The rest-frame UV continuum slope $\beta_{\mathrm{UV}} \approx -2.20 \pm 0.07$ is relatively red for a very young stellar population, suggesting that the UV emission is affected by a small amount of dust attenuation. The full width at half maximum (FWHM) of the intrinsic rest-frame UV light profile is $0.16 \pm 0.01\ \mathrm{arcseconds}$ (corresponding to a de-convolved half-light radius of $260 \pm 20\ \mathrm{parsecs}$). Given its spatial extent, the rest-frame UV emission appears not to be dominated by emission from an active galactic nucleus (AGN). The properties of \mbox{JADES-GS-z14-0} add to the evidence that a population of luminous and massive galaxies is already in place less than $300$ million years after the Big Bang, with number densities more than ten times higher than extrapolations based on pre-JWST observations, as demonstrated by \citetm{Robertson:2024}.

The rest-frame optical nebular emission lines are one of the primary means to characterize the physical conditions in galaxies. However, the redshift of JADES-GS-z14-0 has shifted these lines into the wavelength coverage of MIRI, beyond the wavelength coverage of NIRCam and NIRSpec. The superb performance of JWST \citem{Rigby:2023}, alongside the remarkable brightness of the most extreme high-redshift galaxies, will allow MIRI to explore the rest-frame optical regime and provide important insights into the nature of the earliest galaxies. The first example of MIRI being used at the redshift frontier was with the spectroscopic identification of the rest-frame optical emission lines $\mathrm{[OIII]}\lambda\lambda4959,5007$ and $\mathrm{H}\alpha$ in the galaxy GHZ2/GLASS-z12 at $z = 12.33 \pm 0.02$ \citem{Zavala:2024}. These results  highlight the power of combining observations from MIRI, NIRCam, and NIRSpec to understand the properties of the very first galaxies.

In this work, we present the robust photometric detection of JADES-GS-z14-0 with the Mid-Infrared Instrument (MIRI) at an observed wavelength of $\lambda_{\mathrm{obs}} \approx 7.7\ \mu\mathrm{m}$, corresponding to rest-frame wavelengths of $\lambda_{\mathrm{rest}} \approx 4400-5700\ \Angstrom$. These ultra-deep observations with MIRI/F770W provide more information about the nature of this remarkable galaxy, and are among the deepest mid-infrared integrations to date, with an on-source integration time of $t_{\mathrm{obs}} \approx 23.8\ \mathrm{hr}$ for JADES-GS-z14-0.

\begin{figure*}
    \centering
    \includegraphics[width=1.0\textwidth]{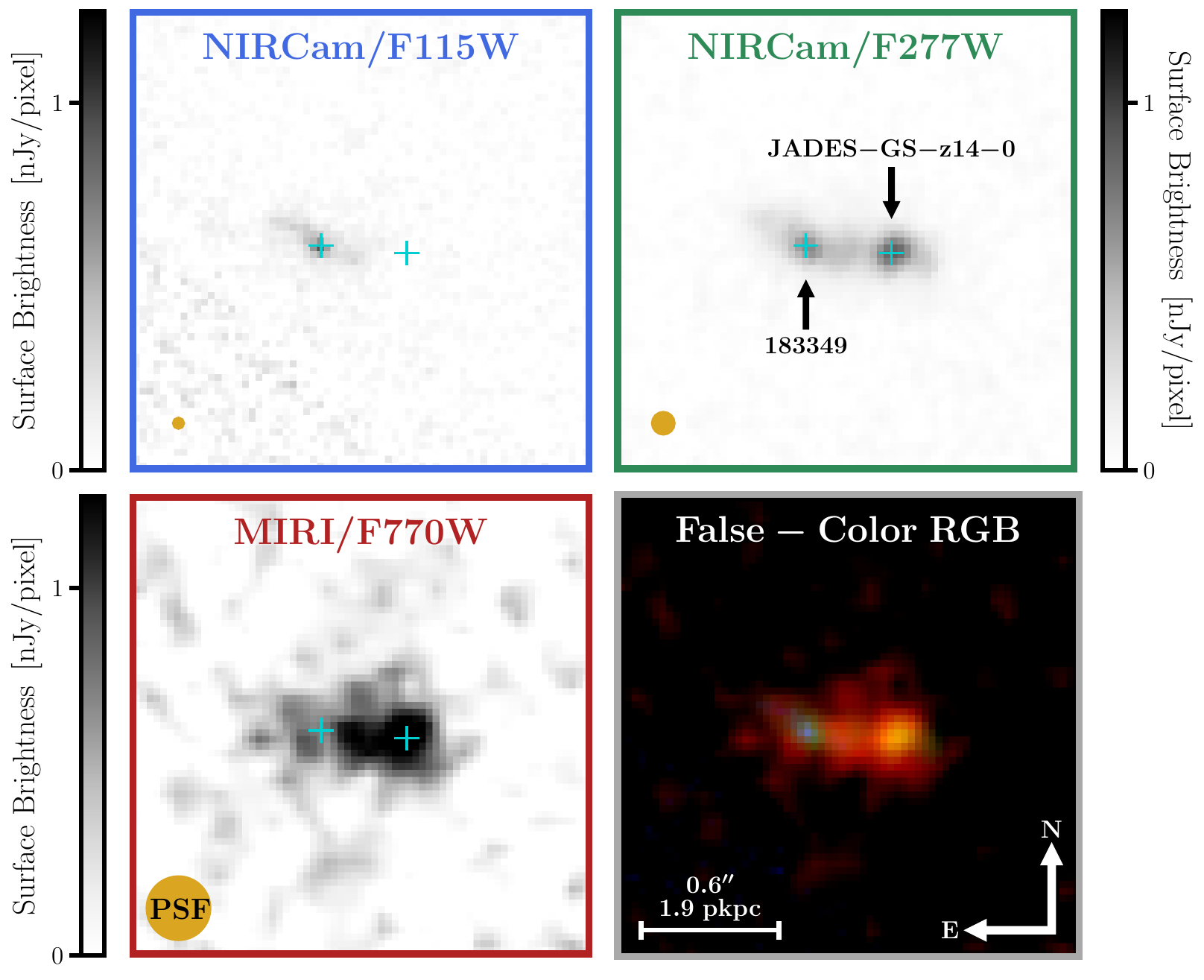}
    \caption{\textbf{A distant galaxy spectroscopically confirmed by the JWST Advanced Deep Extragalactic Survey (JADES).} This galaxy was initially selected from ultra-deep NIRCam and MIRI imaging with JWST (F770W-F277W-F115W shown as an RGB false-color mosaic in the lower right). It was targeted for NIRSpec MSA follow-up observations and has been spectroscopically confirmed at redshift $z = 14.32_{-0.20}^{+0.08}$ \protect\citem{Carniani:2024}. JADES-GS-z14-0 is to the right and the foreground galaxy NIRCam ID $183349$ to the left. The apparent color of JADES-GS-z14-0 is caused by the absorption of the NIRCam/F115W flux by the intervening IGM and the rest-frame optical nebular emission line excess in MIRI/F770W relative to NIRCam/F277W.}
    \label{fig:Figure_01}
\end{figure*}

Figure~\ref{fig:Figure_01} presents in an inset panel the JWST F770W-F277W-F115W false-color image for JADES-GS-z14-0. The apparent color of this galaxy is caused by (1) the complete absorption of emission by the intergalactic medium (IGM) in the F115W filter and (2) the excess rest-frame optical emission in the F770W filter relative to the rest-frame UV emission in the F277W filter. \mbox{JADES-GS-z14-0} is close in projection to a foreground galaxy at a separation of roughly $0.4\ \mathrm{arcseconds}$ to the east, which we refer to as NIRCam ID $183349$ \citem{Carniani:2024}. The lensing magnification caused by $183349$ and another low-redshift neighboring object (roughly $2.2\ \mathrm{arcseconds}$ to the south) is estimated to be small with a lensing magnification factor of $\mu = 1.2$ \citem{Carniani:2024}. All of the analyses and results presented here have been corrected for this magnification factor. 

\begin{table}
    \centering
    \renewcommand{\arraystretch}{1.2}
    \begin{tabular}{l||c}
    \hline
    \hline
    & Photometry for JADES-GS-z14-0 \\
    \hline
    Instrument/\textbf{Filter} & Model Fitting [nJy] \\
    \hline
    NIRCam/\textbf{F090W} & $-2.1 \pm 0.6$ \\
    NIRCam/\textbf{F115W} & $-0.8 \pm 0.4$ \\
    NIRCam/\textbf{F150W} & $1.2 \pm 0.5$ \\
    NIRCam/\textbf{F162M} & $-1.5 \pm 0.9$ \\
    NIRCam/\textbf{F182M} & $13.9 \pm 0.4$ \\
    NIRCam/\textbf{F200W} & $34.8 \pm 0.5$ \\
    NIRCam/\textbf{F210M} & $46.5 \pm 0.6$ \\
    \hline
    NIRCam/\textbf{F250M} & $47.2 \pm 0.5$ \\
    NIRCam/\textbf{F277W} & $55.1 \pm 0.6$ \\
    NIRCam/\textbf{F300M} & $49.8 \pm 0.5$ \\
    NIRCam/\textbf{F335M} & $43.4 \pm 0.5$ \\
    NIRCam/\textbf{F356W} & $47.3 \pm 0.5$ \\
    NIRCam/\textbf{F410M} & $46.1 \pm 0.8$ \\
    NIRCam/\textbf{F444W} & $46.9 \pm 0.6$ \\
    \hline
    MIRI/\textbf{F770W} & $74.4 \pm 5.6$ \\
    \hline
    \end{tabular}
    \caption{\textbf{
    Photometry for JADES-GS-z14-0.} Columns: (1) instrument and filter combinations and (2) fiducial model fitting photometry assuming an extended morphology.}
    \label{tab:Photometry}
\end{table}

Interpreting the ultra-deep MIRI observations in the context of the ultra-deep NIRCam observations requires measuring the flux density in MIRI relative to the flux density in one of the long-wavelength NIRCam filters. The FWHMs of the F770W and F444W point spread functions (PSFs)  are $0.269\ \mathrm{arcseconds}$ and $0.145\ \mathrm{arcseconds}$, respectively, which makes separating JADES-GS-z14-0 from $183349$ challenging, but essential. To meet this challenge, we measure photometry by performing model fitting on the individual exposures prior to mosaicing, illustrated by Extended Data Figure~\ref{fig:ExtendedFigure_01} in the Methods Section. JADES-GS-z14-0 and $183349$ are fit simultaneously in all of the NIRCam and MIRI exposures with \texttt{ForcePho} (B. D. Johnson et al., in preparation) and \texttt{GALFIT} \citem{Peng:2002, Peng:2010}. We model the sources with S\'{e}rsic light profiles. An image of the difference between the observations and our model shows only small residuals, demonstrating the validity of the modeling. We measure the F444W flux density to be $f_{\mathrm{F444W}} = 46.9 \pm 0.6\ \mathrm{nJy}$ and the F770W flux density to be $f_{\mathrm{F770W}} = 74.4 \pm 5.6\ \mathrm{nJy}$ for JADES-GS-z14-0. The measurements correspond to an excess flux of $\Delta f = 27.5 \pm 5.6\ \mathrm{nJy}$ in F770W with respect to F444W. We report these measurements in Table~\ref{tab:Photometry}. Simple PSF photometry confirms these results (see Extended Data Figure~\ref{fig:ExtendedFigure_02}). The quoted photometric uncertainties indicate the signal-to-noise but are likely underestimates of the true errors, since we do not account for systematic uncertainties related to, e.g., photometric calibration, background subtraction, and/or parametric assumptions for the intrinsic light profiles of JADES-GS-z14-0 and $183349$.

\begin{figure*}
    \centering
    \includegraphics[width=0.9\textwidth]{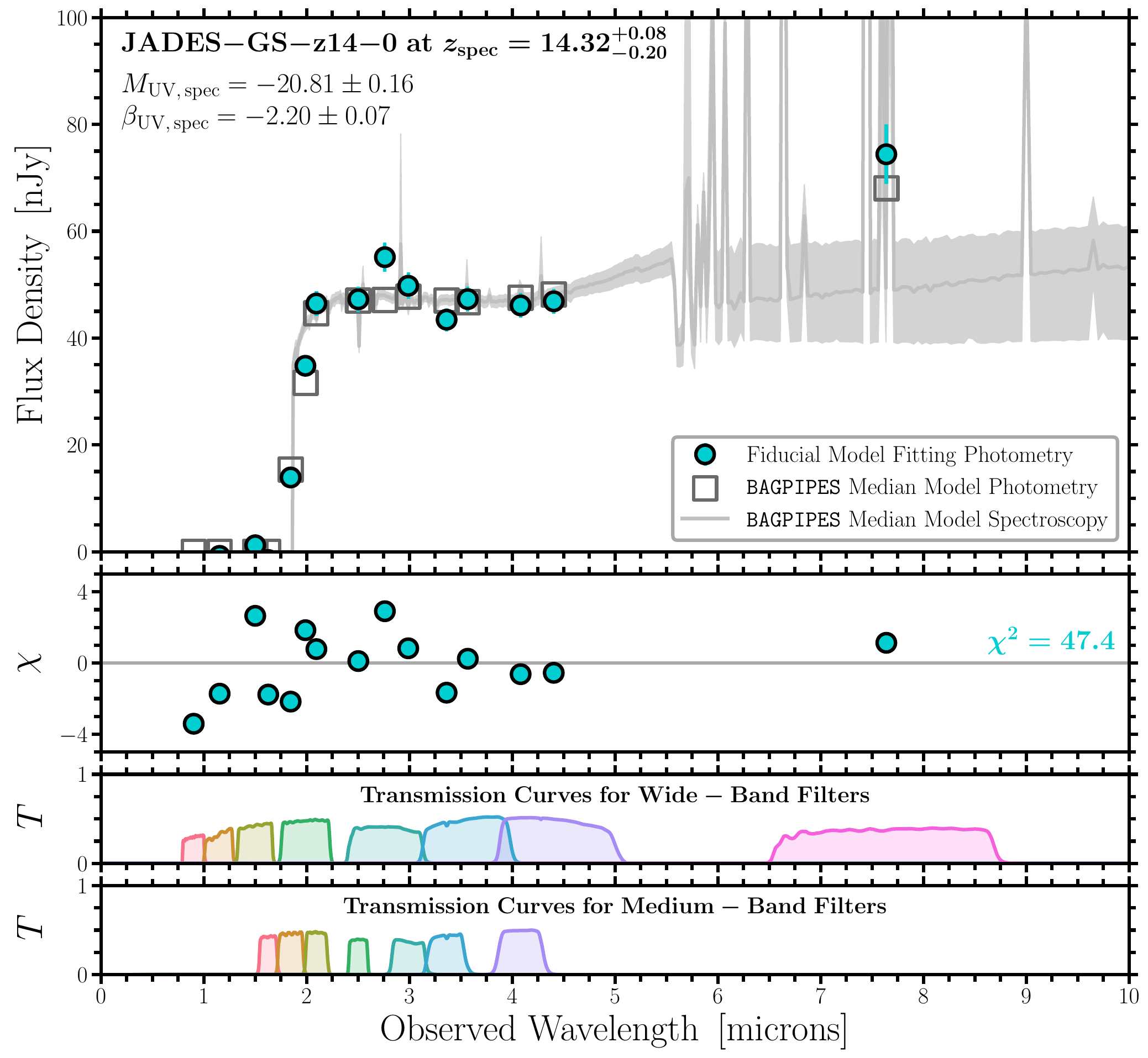}
    \caption{\textbf{Spectral energy distribution (SED) modeling for JADES-GS-z14-0.} In the upper panel, the measured spectral flux density and corresponding uncertainties (fiducial model fitting photometry shown by the blue circles) are used to constrain the various SED models with \texttt{BAGPIPES}. The median of these models is the gray line and unfilled squares, while the $1\sigma$ confidence interval is the shaded region. In the middle panel, the median model photometry is compared with the measured fluxes and uncertainties ($\chi$), while the total $\chi^{2}$ value is reported on the right. In the lower panels, the transmission curves for the various filters are shown. These results suggest that the excess flux in F770W relative to F444W is from nebular emission line contributions, while the underlying continuum is relatively flat at rest-frame optical wavelengths.}
    \label{fig:Figure_02} 
\end{figure*}

To interpret the source of the excess flux in F770W relative to F444W, we model the spectral energy distribution (SED) of JADES-GS-z14-0 using two Bayesian fitting codes: \texttt{BAGPIPES} \citem{Carnall:2018} and \texttt{Prospector} \citem{Johnson:2021}. Generally, we find two types of solutions for the stellar population properties. To explain the red rest-frame UV continuum slope, \texttt{BAGPIPES} prefers solutions with relatively young stellar populations and significant diffuse dust attenuation, while \texttt{Prospector} prefers solutions with older stellar populations but insignificant diffuse dust attenuation. The \texttt{Prospector} models contain a strong Balmer break (see Extended Data Figure~\ref{fig:ExtendedFigure_04}), predicting stellar masses that are nearly an order of magnitude larger than the \texttt{BAGPIPES} models (see Extended Data Table~\ref{tab:Properties_Prospector}), rivaling the maximum expected halo mass for galaxies at these redshifts. Additionally, the \texttt{Prospector} models have the bulk of their stars forming at $z \approx 18-20$ (corresponding to median mass-weighted stellar ages of $t_{\ast} \approx 80-100$ million years), with no recent star formation, which exacerbates concerns about the predicted stellar mass. The lack of recent star formation would be unexpected given the predicted burstiness of star formation in early galaxies \citem{Sun:2023}. \citetm{Wilkins:2023a} show that galaxies at the redshift frontier should have stellar masses, stellar ages, and diffuse dust attenuations that are consistent with the predictions from the \texttt{BAGPIPES} models (see discussion in the Methods Section). Furthermore, \citetm{Wilkins:2024} show that the strength of the Balmer break is reduced substantially with a top-heavy IMF, as is likely for JADES-GS-z14-0 (see discussion below). For these reasons, we consider the solutions from \texttt{BAGPIPES} to be more likely than the solutions from \texttt{Prospector}. Although we are unable to reject them on a formal basis, the \texttt{Prospector} fits would have radical implications for models of early galaxy evolution. 

Figure~\ref{fig:Figure_02} shows the measured photometry in all of the available NIRCam and MIRI filters, alongside the inferred spectral energy distribution (SED) determined with \texttt{BAGPIPES}. This fitting of the SED is self-consistently able to constrain the properties of the (1) stellar populations, (2) dust attenuation, and (3) nebular gas. For simplicity, SEDs are modeled assuming a Kroupa stellar initial mass function (IMF)\citem{Kroupa:2002}. Absorption from the IGM and attenuation from diffuse interstellar dust are included, along with nebular contributions from continuum and emission lines. To understand the effects of differing star-formation histories (SFHs), we ran models for: (1) a parametric constant SFH; (2) a parametric delayed-tau SFH; and (3) a non-parametric continuity SFH. Comparing the measured photometry and corresponding uncertainties (filled circles) with the median of the inferred photometry from the SED modeling (unfilled squares) demonstrates the success in the modeling. The $1\sigma$ confidence interval (shaded regions) suggests that the majority (i.e., at least $50\%$, and up to $100\%$) of the excess flux in the F770W filter is from the nebular emission lines $\mathrm{H}\beta$ and $\mathrm{[OIII]}\lambda\lambda4959,5007$, while the underlying continuum in the rest-frame optical is flat and consistent with the measured flux density in F444W.

\begin{figure*}
    \centering
    \includegraphics[width=1.0\textwidth]{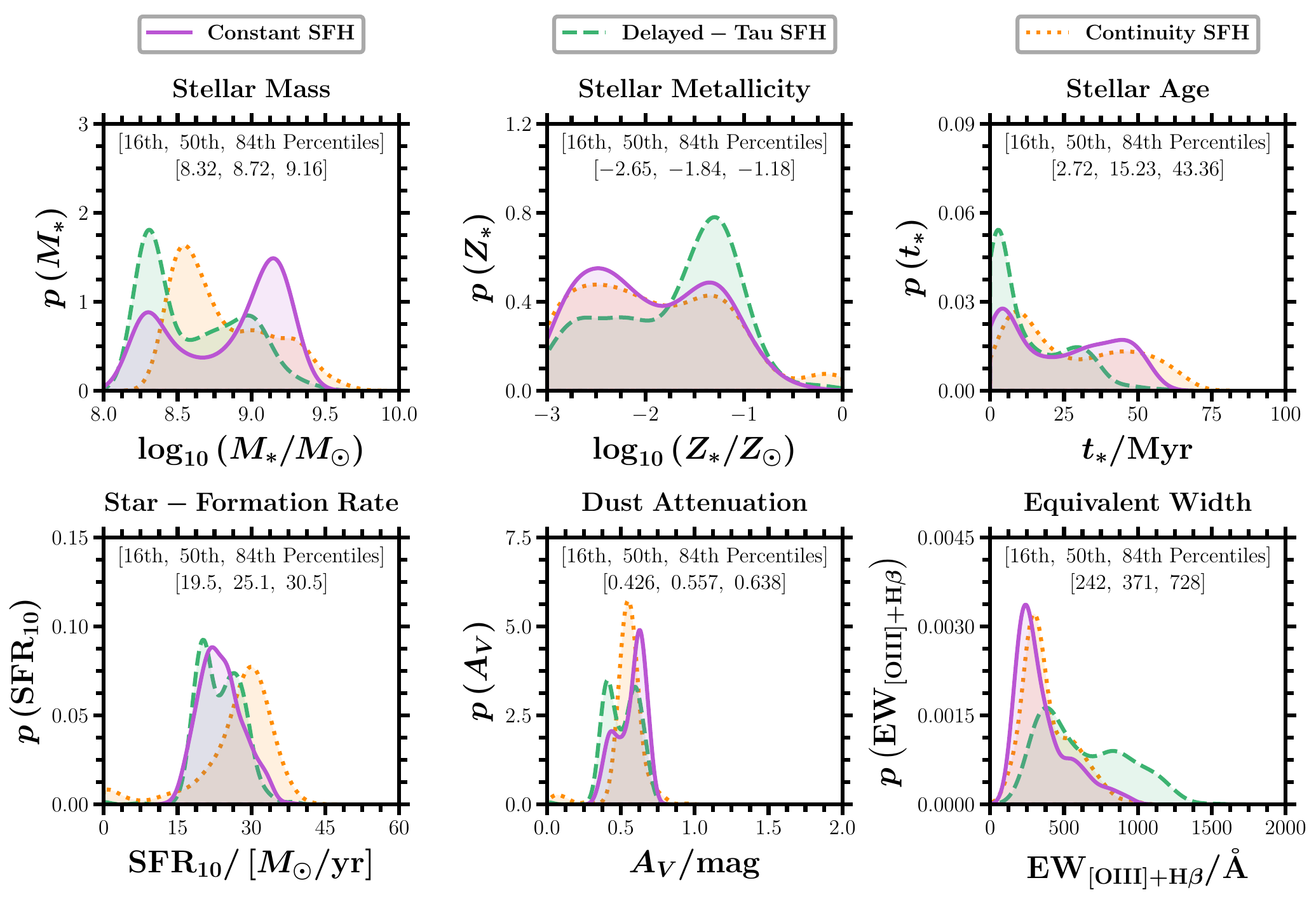}
    \caption{\textbf{Stellar population synthesis modeling using the NIRCam and MIRI photometry.} The measured fluxes and uncertainties are used to constrain the various SED models with \texttt{BAGPIPES}. Shown are the posterior distributions of the stellar mass (\unboldmath$M_{\ast}$ in the upper left), stellar metallicity (\unboldmath$Z_{\ast}$ in the upper middle), mass-weighted stellar age (\unboldmath$t_{\ast}$ in the upper right), star-formation rate averaged over the previous $10$ million years (\unboldmath$\mathrm{SFR}_{10}$ in the lower left), diffuse dust attenuation as measured in the $V$-band (\unboldmath$A_{V}$ in the lower middle), and rest-frame equivalent width of \unboldmath$\mathrm{[OIII]}+\mathrm{H}\beta$ (\unboldmath$\mathrm{EW}_{\mathrm{[OIII]}+\mathrm{H}\beta}$ in the lower right). We report the 16th, 50th, and 84th percentiles after combining the posterior distributions from the various SED models. Three different star-formation histories (SFHs) are assumed: parametric constant SFH (shown by the solid purple lines), parametric delayed-tau SFH (shown by the dashed green lines), and non-parametric continuity SFH (shown by the dotted orange lines).}
    \label{fig:Figure_03}
\end{figure*}

Figure~\ref{fig:Figure_03} and Extended Data Table~\ref{tab:Properties_BAGPIPES} presents the marginalized distributions for the \texttt{BAGPIPES} constraints on stellar mass ($M_{\ast}/M_{\odot}$ in the upper left), stellar metallicity ($Z_{\ast}/Z_{\odot}$ in the upper middle), mass-weighted stellar age ($t_{\ast}/\mathrm{Myr}$ in the upper right), star-formation rate averaged over the previous $10$ million years ($\mathrm{SFR}_{10}/[M_{\odot}/\mathrm{yr}]$ in the lower left), diffuse dust attenuation as measured in the $V$-band ($A_{V}/\mathrm{mag}$ in the lower middle), and rest-frame equivalent width of $\mathrm{[OIII]}+\mathrm{H}\beta$ ($\mathrm{EW}_{\mathrm{[OIII]}+\mathrm{H}\beta}/\mathrm{\mathring{A}}$ in the lower right). The equivalent width is measured for the combined rest-frame optical nebular emission lines $\mathrm{H}\beta$ and $\mathrm{[OIII]}\lambda\lambda4959,5007$. Similarly, Extended Data Figure~\ref{fig:ExtendedFigure_03} in the Methods Section presents constraints on the joint posterior distributions and SFHs with \texttt{BAGPIPES}. The median derived stellar mass $M_{\ast} \approx 10^{8.7}\ M_{\odot}$ (with a $1\sigma$ confidence interval of $M_{\ast} \approx 10^{8.3}-10^{9.2}\ M_{\odot}$) is nearly one-tenth the current value for the Milky Way. The median star-formation rate $\mathrm{SFR} \approx 25\ M_{\odot}/\mathrm{yr}$ (with a $1\sigma$ confidence interval of $\mathrm{SFR} \approx 20-31\ M_{\odot}/\mathrm{yr}$) is consistent with expectations based on the empirical star-forming main sequence derived at lower redshifts ($z \approx 8$)\citem{Popesso:2023}. Taken together with the measured half-light radius, the star-formation rates imply star-formation rate surface densities $\Sigma_{\mathrm{SFR}} \approx 64\ M_{\odot}/\mathrm{yr}/\mathrm{pc}^{2}$ (with a $1\sigma$ confidence interval of $\Sigma_{\mathrm{SFR}} \approx 50-78\ M_{\odot}/\mathrm{yr}/\mathrm{pc}^{2}$), comparable to the most vigorous starbursts observed in the local Universe \citem{Genzel:2010}. 

The inferred stellar masses and star-formation rates are deduced from the emission produced by massive stars and an assumed ``local'' IMF. However, it is likely that the formation of low-mass stars is strongly suppressed both due to the high temperature of the cosmic microwave background ($T \approx 60\ \mathrm{K}$ at $z = 20$)\citem{Chon:2022} and the low metallicity of galaxies at $z > 10$ \citem{Steinhardt:2023, Chon:2024}. Recent work has found that stellar masses can be reduced by factors of three (or more) by changing assumptions about the IMF, without affecting the resulting SED \citem{Woodrum:2024}, which our own analysis confirms.

Returning to the \texttt{BAGPIPES} constraints on physical properties, the median mass-weighted stellar age $t_{\ast} \approx 15$ million years (with a $1\sigma$ confidence interval of $t_{\ast} \approx 3-43$ million years) is consistent with values measured for some of the youngest galaxies observed at lower redshifts ($z \approx 8$)\citem{Endsley:2023, Endsley:2024}. Similarly, the median rest-frame equivalent width of $\mathrm{[OIII]}+\mathrm{H}\beta$ is $\mathrm{EW}_{\mathrm{[OIII]}+\mathrm{H}\beta} \approx 370\ \Angstrom$ (with a $1\sigma$ confidence interval of $\mathrm{EW}_{\mathrm{[OIII]}+\mathrm{H}\beta} \approx 240-730\ \Angstrom$), consistent with values measured for those same galaxies observed at lower redshifts ($z \approx 8$)\citem{Endsley:2023, Endsley:2024}.  The median diffuse dust attenuation $A_{V} \approx 0.56\ \mathrm{AB\ mag}$ (as measured in the $V$-band, with a $1\sigma$ confidence interval of $A_{V} \approx 0.43-0.64\ \mathrm{AB\ mag}$) suggests a small amount of attenuation at rest-frame optical wavelengths. A detailed discussion and interpretation of the dust content for \mbox{JADES-GS-z14-0} is presented in \citetm{Carniani:2024}. Finally, the median stellar metallicity  $Z_{\ast} \approx 0.014\ Z_{\odot}$ (with a $1\sigma$ confidence interval of $Z_{\ast} \approx 0.002-0.066\ Z_{\odot}$) is largely unconstrained but consistent with metallicities that are less than ten percent of the solar value. These results agree with those determined by \citetm{Carniani:2024}, who derive a stellar metallicity from SED modeling with \texttt{BEAGLE} \citem{Chevallard:2016}. Assuming the stellar and gas-phase metallicities are the same, they find $\mathrm{log}_{10}\left(\mathrm{O/H}\right)+12 = 7.2_{-0.4}^{+0.7}$. Our own SED fitting finds an equivalent gas-phase metallicity of $\mathrm{log}_{10}\left(\mathrm{O/H}\right)+12 = 6.9_{-0.8}^{+0.7}$.

The gas-phase metallicity encodes valuable information about the baryonic processes shaping the formation and evolution of galaxies. It has been found to be correlated with the emission line ratio $\mathrm{[OIII]}/\mathrm{H}\beta$, which compares the strengths of the collisionally excited $\mathrm{[OIII]}\lambda\lambda4959,5007$ lines with the Balmer recombination $\mathrm{H}\beta$ line. We can place loose constraints on this emission line ratio by combining the stellar population properties (and the predicted rest-frame optical continuum) with the measured excess flux in F770W relative to F444W. We proceed analogously to the fundamental relations derived by \citetm{Kennicutt:1998} and compare inferred star-formation rates with observed $\mathrm{H}\beta$ line luminosities. The relationship between these two quantities is derived for a sample of galaxies at $z \approx 8$ with $\mathrm{H}\beta$ line flux measurements from NIRSpec/PRISM observations, which are from the JADES Data Release 3 (DR3)\citem{D'Eugenio:2024}. Combining our derived calibration with the inferred star-formation rates from \texttt{BAGPIPES} yields an $\mathrm{H}\beta$ line flux $F_{\mathrm{H}\beta} = 7.9_{-1.8}^{+1.7} \times 10^{-19}\ \mathrm{ergs/s/cm}^{2}$. Comparing this derived quantity with two assumptions about the underlying continuum at rest-frame optical wavelengths yields $\mathrm{[OIII]}/\mathrm{H}\beta = 2.5_{-0.6}^{+0.9}$ and $1.9_{-0.7}^{+2.6}$ (see Extended Data Figure~\ref{fig:ExtendedFigure_05}). For comparison, the typical value is $\mathrm{[OIII]}/\mathrm{H}\beta \approx 6$ for the aforementioned sample of galaxies at $z \approx 8$ from JADES DR3 \citem{D'Eugenio:2024}. These results imply that the measured excess flux at $7.7\ \mu\mathrm{m}$ includes a significant contribution from $\mathrm{[OIII]}\lambda\lambda4959,5007$, but with a gas-phase metallicity that is smaller than typical values at $z \approx 8$ (see discussion in the Methods Section).

The detection of JADES-GS-z14-0 at $z > 14$ by MIRI demonstrates its power in understanding the properties of the earliest galaxies. The most plausible solution for the observed flux at $7.7\ \mu\mathrm{m}$ with MIRI suggests significant contributions from the nebular emission lines $\mathrm{[OIII]}\lambda\lambda4959,5007$, which indicates metal enrichment for this galaxy. An alternative model is possible, but less likely, since it suggests an extreme stellar mass and a strong Balmer break from evolved stars. Deep spectroscopic follow-up observations with MIRI/LRS are required to disentangle these interpretations by directly measuring the contributions to this flux from the nebular emission lines $\mathrm{H}\beta$ and $\mathrm{[OIII]}\lambda\lambda4959,5007$. Such observations would also include $\mathrm{H}\alpha$, producing a direct measurement of the star-formation rate. Given the size, luminosity, and redshift of JADES-GS-z14-0, these measurements would build on a truly unique opportunity to study galaxy formation when the Universe was less than $300$ million years old.

%%%%%%%%%%%%%%%%%%%%%%%%%%%%%%%%%%%%%%%%%%%%%%%%%%

%%%%%%%%%%%%%%%%%%%%% METHODS %%%%%%%%%%%%%%%%%%%%

\clearpage
\makeatletter
\renewcommand{\fnum@table}{Extended Data Table \thetable}
\renewcommand{\fnum@figure}{Extended Data Figure \thefigure}
\makeatother
\setcounter{figure}{0}
\setcounter{table}{0}

\section{Methods}
\label{Section:Methods}

Throughout this work, we report wavelengths in air and adopt the standard flat $\Lambda$CDM cosmology from Planck18 with $H_{0} = 67.4\ \mathrm{km/s/Mpc}$ and $\Omega_{m} = 0.315$\citem[][]{Planck:2020}. All magnitudes are in the AB system\citem{Oke:1983}. Uncertainties are quoted as $68\%$ ($1\sigma$) confidence intervals, unless otherwise stated.

\subsection{Observations}
\label{Subsection:Observations}

The observations used in this work consist of infrared imaging with the Near Infrared Camera (NIRCam) and the Mid-Infrared Instrument (MIRI) in the Great Observatories Origins Deep Survey South (GOODS-S)\citem{Giavalisco:2004} field, near the Hubble Ultra Deep Field (HUDF)\citem{Beckwith:2006} and the JADES Origins Field (JOF)\citem{Eisenstein:2023b}. The NIRCam data were primarily observed as part of the JWST Advanced Deep Extragalactic Survey (JADES)\citem{Eisenstein:2023a}, but also as part of the First Reionization Epoch Spectroscopic COmplete Survey (FRESCO)\citem{Oesch:2023}. The MIRI data were observed as part of JADES.

JADES-GS-z14-0 was observed with NIRCam in four separate programs, which we separate into three categories based on observing time. Ultra-deep observations were conducted with program IDs 1210 (N. L\"{u}tzgendorf) and 3215 (D. Eisenstein) across an area of roughly $10$ square arcminutes using fourteen photometric filters, including seven wide bands (F090W, F115W, F150W, F200W, F277W, F356W, and F444W) and seven medium-bands (F162M, F182M, F210M, F250M, F300M, F335M, and F410M). These were observed October 20 through 24 of 2022 (for program ID 1210) and October 16 through 24 of 2023 (for program ID 3215), with a range in integration times of $39-73\ \mathrm{hr}$ for each of the wide-bands and $55-165\ \mathrm{hr}$ for each of the medium-bands, reaching $5\sigma$ depths of $1.9-3.0$ nJy and $1.6-2.6$ nJy, respectively. Medium-depth observations were conducted with program ID 1180 (D. Eisenstein) across an area of roughly $40$ square arcminutes using eight filters, including the same seven wide bands as program ID 1210 plus one medium-band (F410M). These were observed September 29 through October 5 of 2022 and September 28 through October 3 of 2023, with a range in integration times of $6-8\ \mathrm{hr}$ for each of the filters, reaching $5\sigma$ depths of $4.1-7.3$ nJy. Shallow observations were conducted with program ID 1895 (P. Oesch) across an area of roughly $60$ square arcminutes using three filters (F182M, F210M, and F444W). These were observed November 13 through 18 of 2022, with a range in integration times of $0.25-1.25\ \mathrm{hr}$, reaching $5\sigma$ depths of $10-12$ nJy. Depths are estimated using $0.2$ arcsecond radius circular apertures assuming point-source morphologies. The aforementioned NIRCam observations have been presented and discussed extensively in the literature \citem{Eisenstein:2023a, Eisenstein:2023b, Oesch:2023, Rieke:2023, Robertson:2023, Robertson:2024}.

\begin{figure*}
    \centering
    \includegraphics[width=1.0\textwidth]{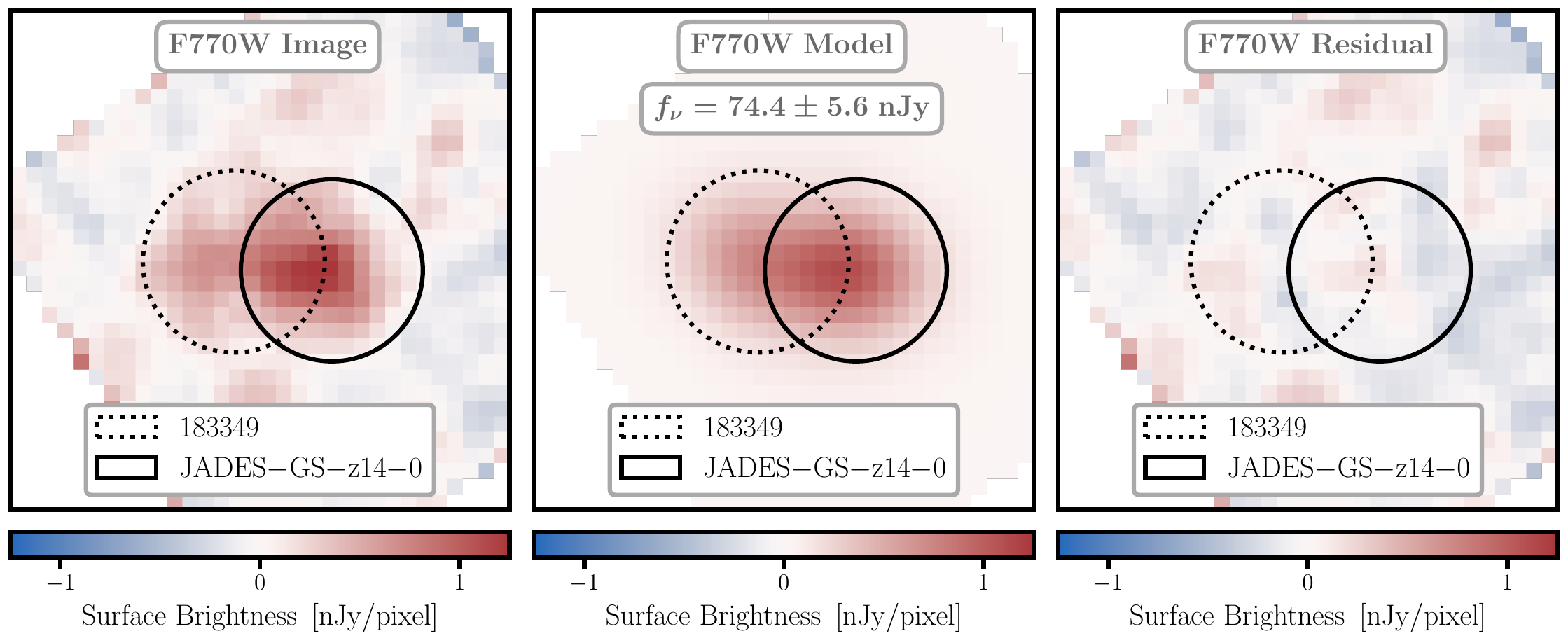}
    \caption{\textbf{An example of the model fitting photometry for JADES-GS-z14-0.} The individual exposures from the MIRI/F770W imaging are modeled assuming an extended morphology. These images are dominated by the diffraction width of the F770W PSF ($\mathrm{FWHM} = 0.269\ \mathrm{arcseconds}$), which is nearly twice the measured half-light angular diameter with NIRCam ($\mathrm{FWHM} = 0.145\ \mathrm{arcseconds}$). We measure photometry by fitting surface brightness profiles to the flux image of the object and its neighbors (left panel), allowing us to construct accurate models (middle panel) which leave only slight residuals (right panel) relative to the data. The thumbnails are $2^{\prime\prime} \times 2^{\prime\prime}$ which corresponds to roughly $6.5\ \mathrm{pkpc} \times 6.5\ \mathrm{pkpc}$ at the observed redshift ($z = 14.32$). We note that the mean stacked flux image of the individual exposures is provided in the left panel for illustrative purposes, rather than the final image mosaic from Figure~\ref{fig:Figure_01} and Extended Data Figure~\ref{fig:ExtendedFigure_02}.}
    \label{fig:ExtendedFigure_01}
\end{figure*}

JADES-GS-z14-0 was also observed with MIRI as a coordinated parallel to NIRCam with program ID 1180 (D. Eisenstein). The MIRI imaging includes four pointings near the JOF at a position angle of $300^{\circ}$, which together produce an ultra-deep contiguous mosaic of roughly $8.8$ square arcminutes. In all four of the pointings, we conducted two separate nine-point dithers of $1361$ second individual exposures with MIRI for five different NIRCam filter pairs ($2 \times 9 \times 5$ total exposures), for a total exposure time of roughly $61.3\ \mathrm{ks}$ with MIRI for each of the two dithers. We conducted two additional four-point dithers of $1361$ second individual exposures with MIRI for three different NIRCam filter pairs ($2 \times 4 \times 3$ total exposures), for a total exposure time of roughly $16.3\ \mathrm{ks}$ seconds for each of the two dithers. The readout mode $\mathrm{SLOWR1}$ was utilized with $57$ groups to minimize data volume. These data were obtained on September 29 through October 5 of 2022 and September 28 through October 3 of 2023, with a typical on-source integration time of roughly $43.1\ \mathrm{hr}$ ($155.2\ \mathrm{ks}$), reaching a $5\sigma$ depth of $21$ nJy ($28.1\ \mathrm{AB\ mag}$) using $0.4$ arcsecond radius circular apertures and assuming point-source morphologies as before. At the location of JADES-GS-z14-0, which falls near the edge of the MIRI imaging, the typical on-source integration time is roughly $23.8\ \mathrm{hr}$ ($85.7\ \mathrm{ks}$) which corresponds to a $5\sigma$ depth of $28$ nJy ($27.8\ \mathrm{AB\ mag}$). These MIRI observations have been presented and discussed previously at half depth in the literature \citem{Alberts:2024}. This is the first presentation and discussion of the full depth MIRI observations, which are some of the deepest images ever taken in the mid-infrared, alongside the ultra-deep MIRI Deep Imaging Survey (MIDIS) results in the HUDF at $5.6\ \mu\mathrm{m}$\citem{Rinaldi:2023, Boogaard:2024}. The observations presented here will be described with more detail in a forthcoming paper from the JADES collaboration (S. Alberts et~al., in preparation).

\subsection{Image Reduction}
\label{Subsection:ImageReduction}

A detailed description of the reduction, mosaicing, source detection, and photometric measurements for the NIRCam data is provided in the first JADES data release in GOODS-S \citem{Rieke:2023} and in \citetm{Carniani:2024}. Similarly, a detailed description of the reduction, mosaicing, source detection, and photometric measurements for the MIRI data will be provided in an upcoming JADES data release in GOODS-S (S. Alberts et~al., in preparation), but has already been partially introduced \citem{Alberts:2024, Lyu:2024, Perez:2024}. We briefly summarize the main steps of the reduction and mosaicing process for the MIRI data. Raw images are processed with the JWST Calibration Pipeline (v1.12.5)\citem{Bushouse_v1p12p5} using the Calibration Reference Data System pipeline mapping 1188, similar to the procedure with NIRCam data. We run Stage~1 of the JWST Calibration Pipeline using all of the default parameters, plus a correction for cosmic ray showers. Stage~2 is run using all of the default parameters, plus an additional custom subtraction of the background using our own super sky backgrounds. Following this step, we perform an additional astrometric correction. Finally, Stage~3 is run using all of the default parameters, but without any further alignment or matching. The final image mosaic is registered to the \textsc{Gaia} DR3 frame \citem{GaiaDR3} and resampled onto the same world coordinate system as the NIRCam image mosaics, but with a $0.060$ arcsecond per pixel grid.

\subsection{Detection and Photometry}
\label{Subsection:DetectionAndPhotometry}

To interpret the ultra-deep MIRI observations in the context of the ultra-deep NIRCam observations, we measure the flux density in MIRI/F770W relative to the flux density in one of the long-wavelength NIRCam filters (i.e., F444W). The MIRI detection is of modest signal-to-noise ($\mathrm{S/N} \approx 13$) with respect to all of the NIRCam detections ($\mathrm{S/N} \approx 25-100$). Furthermore, the MIRI/F770W diffraction-limited PSF (FWHM of $0.269\ \mathrm{arcseconds}$) is larger than NIRCam/F444W (FWHM of $0.145\ \mathrm{arcseconds}$) by nearly a factor of two. JADES-GS-z14-0 and the neighboring foreground galaxy to the east, NIRCam ID $183349$, have similar measured flux densities in F444W and are both morphologically extended (FWHMs of the de-convolved light profiles are both roughly $0.16\ \mathrm{arcseconds}$)\citem{Carniani:2024}. These qualities make the separation (or deblending) of JADES-GS-z14-0 from $183349$ challenging, but essential for physical interpretation of the MIRI observations.

\subsubsection{Detailed Model Fitting Photometry}

Our primary approach for measuring photometry uses detailed model fitting in the vicinity of JADES-GS-z14-0 with the Bayesian fitting code \texttt{ForcePho} (B. D. Johnson et al., in preparation). The motivation for performing these more complicated photometric measurements is the complexity of the region surrounding JADES-GS-z14-0, with multiple bright foreground galaxies within a radius of a few arcseconds. It is important to properly disentangle the relative flux contributions from these galaxies. Furthermore, this region of the sky is located near the edge of the MIRI exposures, which may affect measurements of the photometric uncertainties. We briefly summarize here the main steps in measuring the detailed model fitting photometry. Since the galaxies are better-resolved with NIRCam, we determine their morphological properties (i.e., half-light radius, S\'{e}rsic index, axis ratio, position angle, and total flux for each NIRCam filter) by fitting the NIRCam data and then use the results to constrain the MIRI photometry.

The first step for the NIRCam data is to construct $2.5$ arcsecond by $2.5$ arcsecond cutouts from all of the available exposures, with each of these cutouts centered on the source location of \mbox{JADES-GS-z14-0} from the original photometric catalog. Cutouts are performed on exposures after reduction but prior to mosaicing. All of the detected sources within these cutouts are simultaneously modeled to determine the relative flux contributions. These models assume an intrinsic S\'{e}rsic profile that is identical for each of the different NIRCam filters \citem{Sersic:1968}. We convolve these intrinsic light profiles with Gaussian mixture approximations to the PSF in each of the NIRCam exposures to produce observed light profiles. We utilize the results of this detailed model fitting approach by taking the mean and standard deviation of the resulting total flux posterior distributions, which are sampled with Markov Chain Monte Carlo (MCMC) techniques and account for the covariance between parameters.

JADES-GS-z14-0 and $183349$ are significantly detected (signal-to-noise ratio $\mathrm{S/N} > 5$) in the vast majority of the individual NIRCam exposures, but they are only marginally detected ($\mathrm{S/N} = 2-3$) in most of the individual MIRI exposures. This necessitates different detailed model fitting photometric procedures for NIRCam and MIRI. Extended Data Figure~\ref{fig:ExtendedFigure_01} illustrates the photometric modeling process used to measure the flux density in F770W. The first step is once again to construct $2.5$ arcsecond by $2.5$ arcsecond cutouts, now with all of the available MIRI exposures. All of the detected sources within these cutouts are simultaneously modeled by adopting the inferred morphological properties from the \texttt{ForcePho} fitting to the NIRCam data (e.g., half-light radius, S\'{e}rsic index, axis ratio, and position angle). For JADES-GS-z14-0, the inferred morphological parameters are: half-light radius $r_{1/2} = 0.0788 \pm 0.0006\ \mathrm{arcseconds}$, S\'{e}rsic index $n = 0.877 \pm 0.027$, axis ratio $b/a = 0.425 \pm 0.008$, and position angle $\mathrm{PA} = 1.427 \pm 0.008\ \mathrm{radians}$. For $183349$, the inferred parameters are: $r_{1/2} = 0.0872 \pm 0.0006\ \mathrm{arcseconds}$, $n = 0.815 \pm 0.020$, $b/a = 0.378 \pm 0.006$, and $\mathrm{PA} = 4.171 \pm 0.003\ \mathrm{radians}$. The adopted S\'{e}rsic profile is used to forward-model the total flux in the F770W filter with \texttt{GALFIT} \citem{Peng:2002, Peng:2010}. We adopt the mean and standard deviation of the resulting total flux posterior distributions as fiducial for MIRI. For JADES-GS-z14-0, we measure the F444W flux density to be $f_{\mathrm{F444W}} = 46.9 \pm 0.6\ \mathrm{nJy}$ and the F770W flux density to be $f_{\mathrm{F770W}} = 74.4 \pm 5.6\ \mathrm{nJy}$, corresponding to an excess flux of $\Delta f = 27.5 \pm 5.6\ \mathrm{nJy}$ in F770W with respect to F444W. We report these measurements in Table~\ref{tab:Photometry}. The uncertainties are additionally validated using two different methodologies, producing consistent results within $10\%$ of one another: (1) bootstrapping individual exposures and (2) measuring photometry in regions of the sky that are observed to be empty in the much deeper NIRCam images. For $183349$, we measure the F444W flux density to be $f_{\mathrm{F444W}} = 48.9 \pm 0.6\ \mathrm{nJy}$ and the F770W flux density to be $f_{\mathrm{F770W}} = 46.3 \pm 4.6\ \mathrm{nJy}$.

\begin{figure*}
    \centering
    \includegraphics[width=1.0\textwidth]{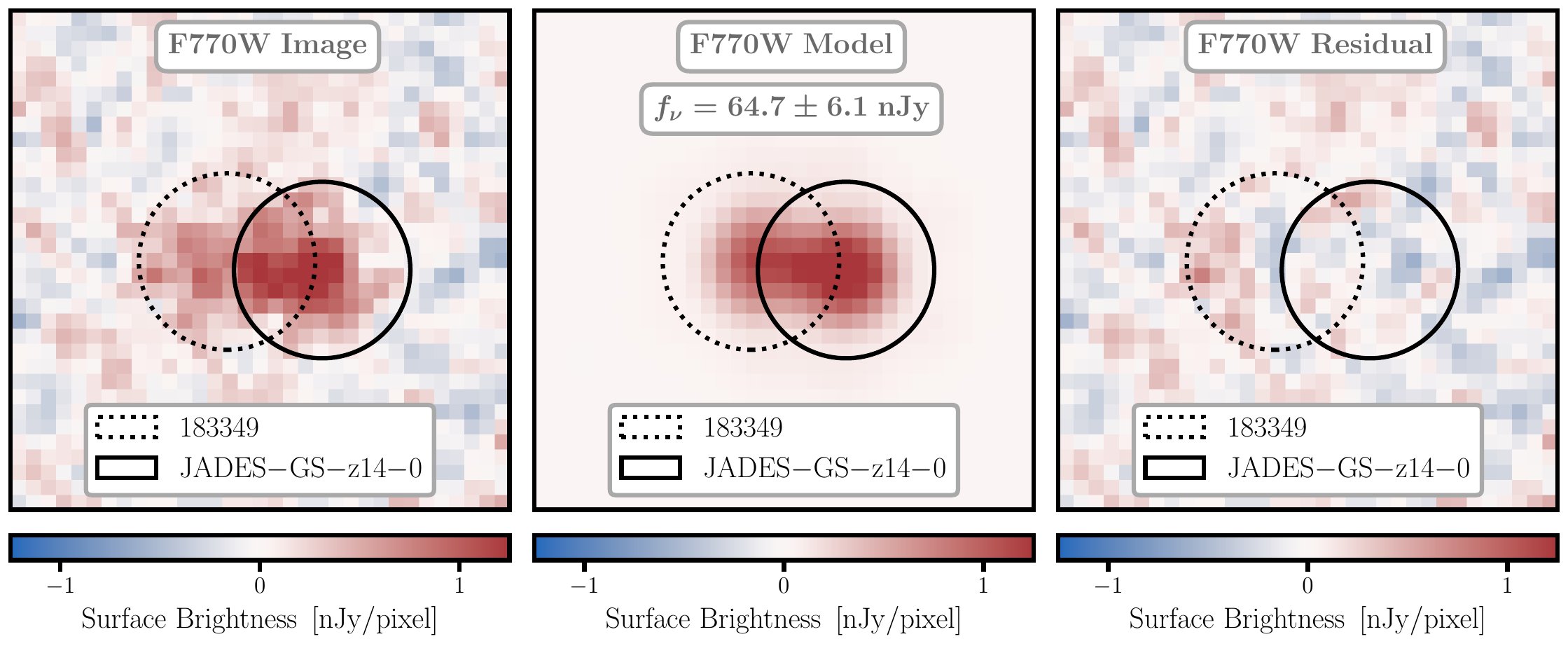}
    \caption{\textbf{An example of the PSF photometry for JADES-GS-z14-0.} The final image mosaic for the MIRI/F770W imaging is modeled assuming a point source morphology for both galaxies. We measure photometry by fitting surface brightness profiles to the flux image of the object and its neighbors (left panel), allowing us to construct an accurate model of the image (middle panel) that leaves only slight residuals (right panel) relative to the data. The thumbnails are $2^{\prime\prime} \times 2^{\prime\prime}$ which corresponds to roughly $6.5\ \mathrm{pkpc} \times 6.5\ \mathrm{pkpc}$ at the observed redshift.}
    \label{fig:ExtendedFigure_02}
\end{figure*}

The model fitting approach assumes that JADES-GS-z14-0 and $183349$ are well described by single S\'{e}rsic light profiles. To explore the impact of this assumption, we additionally  adopt a non-parametric approach. We construct galaxy templates by de-convolving the NIRCam images in the F150W and F444W filters, where the F150W image corresponds to $183349$ alone, while the F444W image corresponds to both JADES-GS-z14-0 and $183349$. We conduct the de-convolution using the Wiener-Hunt method and Gaussian mixture approximations for the PSFs to mitigate noise amplification at high frequencies. We then convolve the templates into the MIRI F770W band and fit the amplitudes of the two templates simultaneously, where we measure on individual exposures and analytically derive the best-estimated values and covariances as before. We assign zero weight to regions below a significance of $3\sigma$ to prevent over-fitting. Finally, we convert the template amplitude to galaxy fluxes according to the \texttt{ForcePho} photometry results in the NIRCam F150W and F444W bands. This method yields a F770W flux density of $f_{\mathrm{F770W}} = 69.3 \pm 6.1\ \mathrm{nJy}$ for JADES-GS-z14-0, which is consistent with the parametric model fitting photometry. Moreover, we apply this non-parametric approach to measure fluxes across all of the NIRCam filters, obtaining results consistent with the reported photometry in Table~\ref{tab:Photometry}. For $183349$, this method yields a F770W flux density of $f_{\mathrm{F770W}} = 47.8 \pm 4.0\ \mathrm{nJy}$.

\subsubsection{PSF Photometry}

Our primary measurements are based on the model fitting approach described above. Alternatively, we could treat both galaxies as point sources, since their measured half-light angular sizes are significantly smaller than the F770W PSF. Therefore, as a check on the model fitting result, we also perform PSF photometry. To measure the observed flux density in F770W relative to F444W, we first convolve the F444W mosaic to the PSF of the F770W filter, which involves convolving the F444W mosaic with a kernel of the difference between the F444W and F770W PSFs. The adopted MIRI PSF is empirically measured from the final image mosaic and accounts for the ``cruciform'' detector artifact \citem{Gaspar:2021}. We simultaneously fit for the flux densities of JADES-GS-z14-0 and $183349$ in both the convolved F444W and unconvolved F770W image mosaics. Extended Data Figure~\ref{fig:ExtendedFigure_02} illustrates the photometric modeling process used to measure the flux density in F770W. The residual images (shown in the right panel) illustrate the validity of the point source assumption and the success in the photometric modeling. For JADES-GS-z14-0, we measure the F444W flux density to be $f_{\mathrm{F444W}} = 46.4 \pm 1.2\ \mathrm{nJy}$ and the F770W flux density to be $f_{\mathrm{F770W}} = 64.7 \pm 6.1\ \mathrm{nJy}$, corresponding to an excess flux of $\Delta f = 18.3 \pm 6.2\ \mathrm{nJy}$ in F770W with respect to F444W. The uncertainty on the F770W flux density is found by injecting fake point sources into the final image mosaic, extracting them with procedures identical to those used for \mbox{JADES-GS-z14-0}, and then calculating the sigma-clipped standard deviation of the difference between the recovered and injected flux densities. These results are consistent with those obtained by both the parametric and non-parametric model fitting approaches. The measured F770W flux density is slightly smaller for the PSF photometry, consistent with a small amount (roughly $10\%$) of galaxy light falling far enough outside the image cores that it is not included in this photometric approach.

\subsubsection{Conclusion}

To summarize, we obtain consistent results for the NIRCam and MIRI photometry of JADES-GS-z14-0 using three different methods. The PSF fitting approach provides a simple baseline that is free of complex modeling assumptions. The model fitting approach adopts both parametric and non-parametric galaxy models while measuring photometry from the individual exposures. This method more accurately accounts for the extended morphologies of JADES-GS-z14-0 and $183349$, which improves measurements of the diffuse galaxy flux. This extended flux is the source of the larger F770W flux densities that are measured with the model fitting approach when compared to the simpler PSF fitting approach.

\begin{table*}
    \centering
    \renewcommand{\arraystretch}{1.4}
    \begin{tabular}{l||c|c|c}
    \hline
    \hline
    & \multicolumn{3}{c}{Assumed Physical Model with \texttt{BAGPIPES}} \\
    \hline
    Inferred Property & Constant SFH & Delayed-Tau SFH & Continuity SFH \\
    \hline
    $\mathrm{log}_{10}\left( M_{\ast}/M_{\odot} \right)$ & ${9.0}_{-0.7}^{+0.2}$ & ${8.5}_{-0.2}^{+0.5}$ & ${8.7}_{-0.2}^{+0.5}$ \\
    $\mathrm{log}_{10}\left( Z_{\ast}/Z_{\odot} \right)$ & ${-2.0}_{-0.7}^{+0.8}$ & ${-1.5}_{-1.0}^{+0.4}$ & ${-2.0}_{-0.7}^{+0.8}$ \\
    $t_{\ast}/\mathrm{Myr}$ & ${22.0}_{-19.2}^{+23.5}$ & ${6.1}_{-4.6}^{+22.6}$ & ${22.1}_{-15.8}^{+29.0}$ \\
    $\mathrm{SFR}_{10}/\left[ M_{\odot}/\mathrm{yr} \right]$ & ${23.4}_{-3.8}^{+5.1}$ & ${23.5}_{-4.2}^{+4.7}$ & ${28.9}_{-7.7}^{+4.0}$ \\
    $\Sigma_{\mathrm{SFR}_{10}}/\left[ M_{\odot}/\mathrm{yr}/\mathrm{pc}^{2} \right]$ & ${59.6}_{-9.8}^{+12.9}$ & ${59.8}_{-10.6}^{+12.0}$ & ${73.5}_{-19.6}^{+10.3}$ \\
    $\mathrm{sSFR}_{10}/\mathrm{Gyr}^{-1}$ & ${26.9}_{-13.2}^{+74.8}$ & ${86.4}_{-60.9}^{+15.2}$ & ${58.5}_{-45.1}^{+44.2}$ \\
    $A_{V}/\left[\mathrm{AB\ mag}\right]$ & ${0.60}_{-0.15}^{+0.06}$ & ${0.52}_{-0.12}^{+0.11}$ & ${0.55}_{-0.06}^{+0.06}$ \\
    $\mathrm{log}_{10}\left( U \right)$ & ${-2.6}_{-0.8}^{+0.5}$ & ${-2.5}_{-0.6}^{+0.3}$ & ${-2.5}_{-0.8}^{+0.4}$ \\
    $\mathrm{EW}_{\mathrm{[OIII]}+\mathrm{H}\beta}/\Angstrom$ & ${284}_{-83}^{+260}$ & ${566}_{-226}^{+399}$ & ${338}_{-90}^{+239}$ \\
    \hline
    \end{tabular}
    \caption{\textbf{Inferred physical properties of JADES-GS-z14-0 from SED modeling with \texttt{BAGPIPES}.} Columns: (1) name and units of the inferred property, (2) measurements using the constant SFH model, (3) measurements using the delayed-tau SFH model, and (4) measurements using the continuity SFH model. Reported measurements are median values and $1\sigma$ confidence intervals.}
    \label{tab:Properties_BAGPIPES}
\end{table*}

\subsection{Stellar Population Synthesis Modeling}
\label{Subsection:StellarPopulationSynthesisModeling}

The combination of ultra-deep MIRI, NIRCam, and NIRSpec observations provides an unprecedented opportunity to study the physical properties of JADES-GS-z14-0. We utilize the Bayesian SED fitting code \texttt{BAGPIPES} (Bayesian Analysis of Galaxies for Physical Inference and Parameter EStimation)\citem{Carnall:2018} to self-consistently model the properties of the stellar populations, dust attenuation, and nebular gas. We choose to sample the posterior distributions of these derived properties with the importance nested sampling code \texttt{nautilus} \citem{Lange:2023}, assuming an effective sample size of $10^4$. Fits are performed on the fiducial model fitting photometry after imposing an error floor of 5\%. Such an error floor is imposed since the quoted photometric uncertainties are likely underestimates of the true errors. This is because we do not account for systematic uncertainties related to, e.g., imperfect photometric calibration, background subtraction, and/or parametric assumptions about the intrinsic light profiles (where we assume a single S\'{e}rsic profile for each source). We briefly summarize here the various components of the assumed physical model.

Stellar populations are derived using pre-defined stellar population synthesis (SPS) models \citem{Chevallard:2016}, which are the 2016 updated version of previous models \citem{Bruzual:2003}, and are determined for a grid of simple stellar-population (SSP) models with various ages and metallicities. We adopt the stellar library from the Medium-resolution Isaac Newton Telescope Library of Empirical Spectra (MILES)\citem{Falcon-Barroso:2011}, in addition to the stellar evolutionary tracks and isochrones from the PAdova and TRieste Stellar Evolution Code (PARSEC)\citem{Bressan:2012}. We assume a Kroupa initial mass function (IMF)\citem{Kroupa:2002} with a lower bound of  $0.08\ M_{\odot}$ and an upper bound of $120\ M_{\odot}$. Absorption from the intergalactic medium (IGM) is modeled after \citetm{Inoue:2014}, which is an updated version of the model from \citetm{Madau:1995}. Dust attenuation is modeled after \citetm{Calzetti:2000} with one free parameter: the diffuse dust attenuation as measured in the $V$-band ($A_{V}$, assuming a uniform prior with $\mathrm{min} = 0.0$ and $\mathrm{max} = 2.0$). Nebular emission (from both emission lines and continuum) is modeled after \citetm{Byler:2017} using the photoionization code \texttt{Cloudy} \citem{Ferland:2013} with one free parameter: the ionization parameter ($\mathrm{log}_{10}[U]$, assuming a uniform prior with $\mathrm{min} = -4.0$ and $\mathrm{max} = -2.0$). The gas-phase metallicity is fixed to the value of the stellar metallicity.

To explore the impact of assuming different star-formation histories (SFHs) on the results, we model the SEDs with two parametric SFHs (constant model and delayed-tau model) and one non-parametric SFH (continuity model). Each of the assumed SFHs has at least two free parameters: the total stellar mass formed ($\mathrm{log}_{10}[M_{\ast}/M_{\odot}]$, assuming a uniform prior with $\mathrm{min} = +6.0$ and $\mathrm{max} = +12.0$) and the stellar metallicity ($\mathrm{log}_{10}[Z_{\ast}/Z_{\odot}]$, assuming a uniform prior with $\mathrm{min} = -3.0$ and $\mathrm{max} = +0.0$). The constant SFH model has one additional free parameter: the galaxy age ($t/\mathrm{Myr}$, assuming a uniform prior with $\mathrm{min} = 1.0$ and $\mathrm{max} = t_{\mathrm{univ}}/\mathrm{Myr}$, where $t_{\mathrm{univ}}$ is the age of the Universe measured with respect to the formation redshift $z_{\mathrm{form}} = 20$). The delayed-tau SFH model has two additional free parameters: the galaxy age ($t/\mathrm{Myr}$, assuming a uniform prior with $\mathrm{min} = 1.0$ and $\mathrm{max} = t_{\mathrm{univ}}/\mathrm{Myr}$) and the $e$-folding time for the delayed-tau component ($\tau/\mathrm{Gyr}$, assuming a log-uniform prior with $\mathrm{min} = 0.001$ and $\mathrm{max} = 30.0$). Finally, the continuity SFH model has four additional free parameters, corresponding to the logarithm of the ratio of the SFRs in the five adjacent time-bins ($R_{\mathrm{SFR}}$, assuming Student's $t$-distribution prior with $\mu = 0.0, \sigma = 0.3$). These time-bins are spaced at lookback times of $0-3$, $3-10$, $10-30$, $30-100$, and $100-t_{\mathrm{univ}}$ million years, which assumes the SFH starts at the formation redshift $z_{\mathrm{form}} = 20$. These physical models have between five and eight free parameters, which should be compared to the eleven photometric detections that we have for JADES-GS-z14-0 (F182M, F200W, F210M, F250M, F277W, F300M, F335M, F356W, F410M, F444W, and F770W).

Figure~\ref{fig:Figure_03} presents the marginalized distributions for the \mbox{\texttt{BAGPIPES}} constraints on stellar mass ($M_{\ast}$), stellar metallicity ($Z_{\ast}$), mass-weighted stellar age ($t_{\ast}$), star-formation rate averaged over the previous $10$ million years ($\mathrm{SFR}_{10}$), diffuse dust attenuation as measured in the $V$-band ($A_{V}$), and rest-frame equivalent width for the nebular emission lines $\mathrm{H}\beta$ and $\mathrm{[OIII]}\lambda\lambda4959,5007$ ($\mathrm{EW}_{\mathrm{[OIII]}+\mathrm{H}\beta}$). Results from each of the assumed SFHs are shown: the parametric constant SFH model is represented by the purple solid lines, the parametric delayed-tau SFH model by the green dashed lines, and the non-parametric continuity SFH model by the orange dotted lines. Throughout, we quote median derived properties and $1\sigma$ confidence intervals by concatenating the results from each of these SFHs. These measurements are reported in Supplementary Table~\ref{tab:Properties_BAGPIPES}, alongside measurements of the star-formation rate surface density ($\Sigma_{\mathrm{SFR}_{10}}$), specific star-formation rate ($\mathrm{sSFR}_{10}$), and ionization parameter ($\mathrm{log}_{10}\left[U\right]$).

\begin{figure*}
    \centering
    \includegraphics[width=1.0\textwidth]{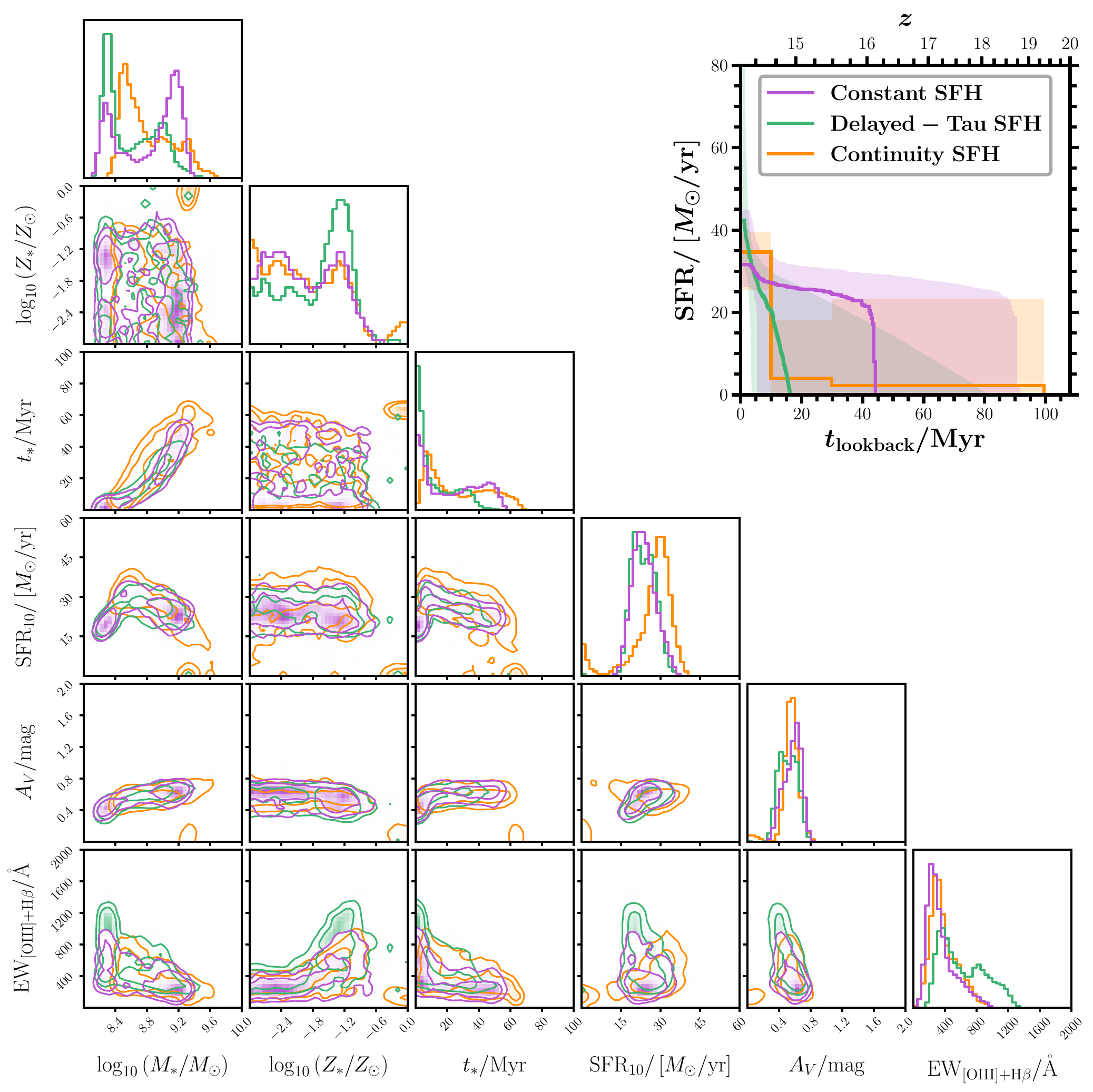}
    \caption{\textbf{Constraints on joint posterior distributions and star formation histories (SFHs) with \texttt{BAGPIPES}.} Various SED models with different SFH assumptions (constant in purple, delayed-tau in green, and continuity in orange) are implemented using \texttt{BAGPIPES} and the measured model fitting photometry. In the lower left, the joint posterior distributions are shown for stellar mass ($M_{\ast}$), stellar metallicity ($Z_{\ast}$), mass-weighted stellar age ($t_{\ast}$), star-formation rate averaged over the previous $10$ million years ($\mathrm{SFR}_{10}$), diffuse dust attenuation as measured in the $V$-band ($A_{V}$), and rest-frame equivalent width for the nebular emission lines $\mathrm{H}\beta$ and $\mathrm{[OIII]}\lambda\lambda4959,5007$ ($\mathrm{EW}_{\mathrm{[OIII]}+\mathrm{H}\beta}$). In the upper right, the derived SFHs are shown. This figure highlights degeneracies between some of the inferred physical parameters.}
    \label{fig:ExtendedFigure_03}
\end{figure*}

Similar to Figure~\ref{fig:Figure_03}, Extended Data Figure~\ref{fig:ExtendedFigure_03} shows the joint posterior distributions for some of the inferred physical parameters in the lower left, alongside the derived SFHs in the upper right. Results from each of the assumed SFHs are provided: the parametric constant SFH model is shown in purple, the parametric delayed-tau SFH model in green, and the non-parametric continuity SFH model in orange. The inferred stellar masses have two peaks in their posterior distributions, one at low-mass ($M_{\ast} \approx 10^{8.5}\ M_{\odot}$) and one at high-mass ($M_{\ast} \approx 10^{9.0}\ M_{\odot}$). The low-mass solutions suggest less stellar continuum at rest-frame optical wavelengths, and therefore larger equivalent widths of the rest-frame optical nebular emission lines $\mathrm{H}\beta$ and $\mathrm{[OIII]}\lambda\lambda4959,5007$. The opposite is true for the high-mass solutions, where the equivalent widths are smaller due to more stellar continuum in the rest-frame optical. The delayed-tau and continuity SFHs prefer the low-mass solution, while the constant SFH finds equal weight to these two solutions. For all of the models, stellar mass is degenerate with stellar age, where the low-mass solution corresponds to younger stellar populations (with mass-weighted ages of a few million years), while the high-mass solution corresponds to older stellar populations (with ages of a few tens of million years). Some of the models suggest more extended periods of star-formation out to lookback times of up to $100$ million years, which would be enough time to enrich this galaxy via Type II supernovae. Despite these differences, the rest of the inferred physical parameters are similar for the two stellar mass and stellar age solutions. All of the inferred physical parameters are fairly well constrained by the existing observations, except for the stellar metallicity (and therefore also the gas-phase metallicity). The results of our SED modeling agree within the quoted uncertainties with the SED modeling by \citetm{Carniani:2024}, where they include JWST/NIRSpec data and adopt various physical models with \texttt{BEAGLE} \citem{Chevallard:2016}.

To explore the impact of adopting different fitting codes, we additionally utilize the Bayesian SED fitting code \texttt{Prospector} (v1.2.0)\citem{Johnson:2021} to self-consistently model the properties of the stellar populations, dust attenuation, and nebular gas. We choose to sample the posterior distributions of these derived properties with the dynamic nested sampling code \texttt{dynesty} (v1.2.3)\citem{Speagle:2020}, assuming an effective sample size of $10^4$. Fits are performed on the model fitting photometry after imposing an error floor of 5\%. The assumed physical model closely follows that of the \texttt{BAGPIPES} modeling, and we briefly summarize here the various components of this model.

Stellar populations are derived with the Flexible Stellar Population Synthesis code (\texttt{FSPS})\citem{Conroy:2009, Conroy:2010} which is accessed through the \texttt{python-FSPS} bindings \citem{Foreman-Mackey:2014}. Stellar evolution is computed by the Modules for Experiments in Stellar Astrophysics package (MESA)\citem{Paxton:2011, Paxton:2013, Paxton:2015, Paxton:2018} while using the synthetic models from MESA Isochrones and Stellar Tracks (MIST)\citem{Dotter:2016, Choi:2016}. The stellar libraries, in addition to the stellar evolutionary tracks and isochrones, are the primary differences between the physical models that we assumed for \texttt{BAGPIPES} and \texttt{Prospector}. Thus, we attribute any differences in the derived physical properties to these differences in the stellar libraries, stellar evolutionary tracks, and isochrones. We assume a Kroupa initial mass function (IMF)\citem{Kroupa:2002} with a lower bound of $0.08\ M_{\odot}$ and an upper bound of $120\ M_{\odot}$. Absorption from the intergalactic medium (IGM) is modeled after \citetm{Madau:1995}. Dust attenuation is modeled after \citetm{Calzetti:2000} with one free parameter: the diffuse dust attenuation as measured in the $V$-band ($A_{V}$, assuming a uniform prior with $\mathrm{min} = 0.0$ and $\mathrm{max} = 2.0$). Nebular emission (from both emission lines and continuum) is modeled after \citetm{Byler:2017} using the photoionization code \texttt{Cloudy} \citem{Ferland:2013} with one free parameter: the ionization parameter ($\mathrm{log}_{10}[U]$, assuming a uniform prior with $\mathrm{min} = -4.0$ and $\mathrm{max} = -1.0$). The gas-phase metallicity is fixed to the value of the stellar metallicity.

\begin{figure*}
    \centering
    \includegraphics[width=0.9\textwidth]{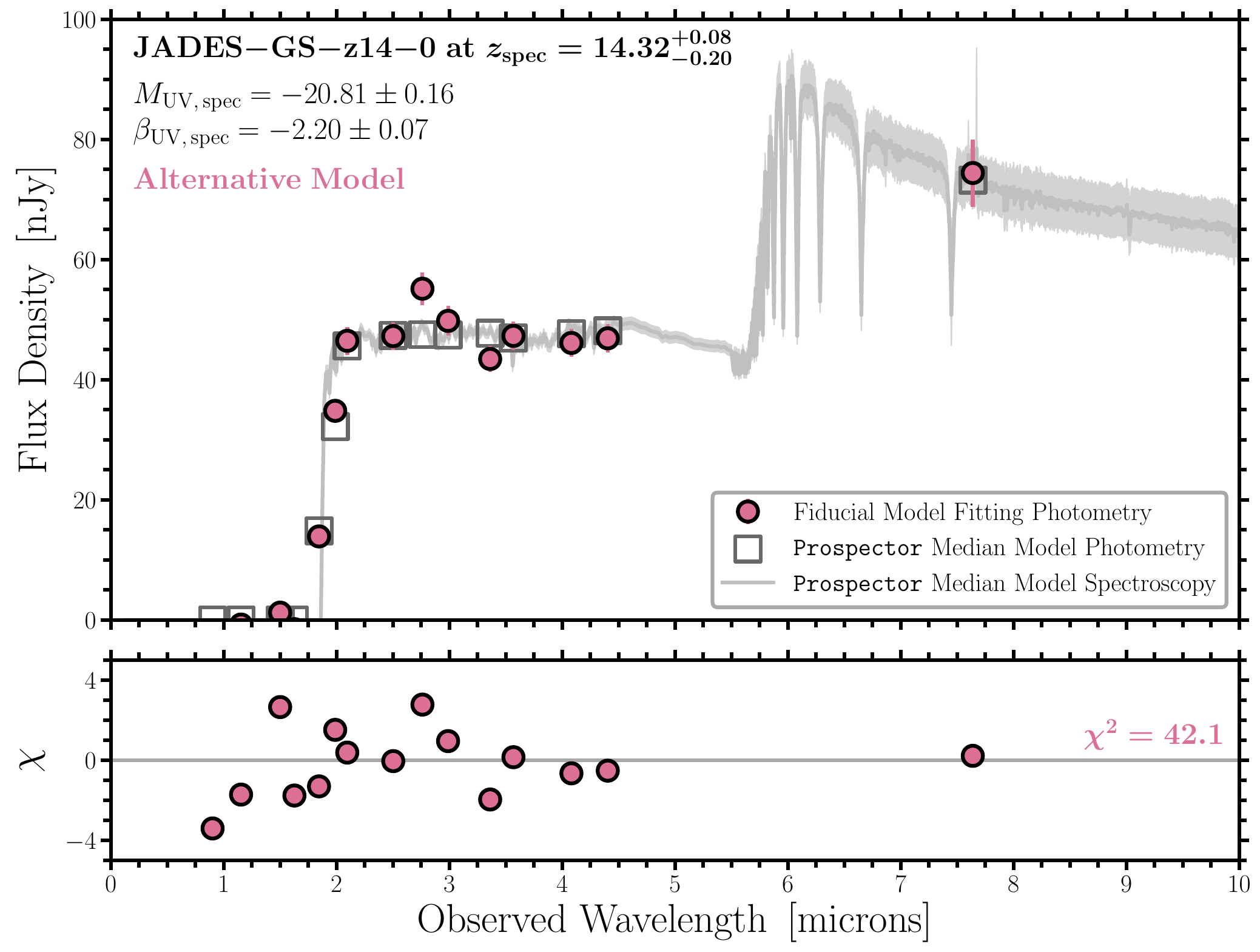}
    \caption{\textbf{SED modeling for JADES-GS-z14-0.} Similar to Figure~\ref{fig:Figure_02}, the measured spectral flux density and corresponding uncertainties are used to constrain the various SED models with \texttt{Prospector}. These results suggest that the excess flux in F770W relative to F444W is from stellar continuum rather than nebular emission line contributions, producing strong Balmer breaks in the  models, which we find unlikely given the extreme inferred stellar masses and stellar ages.}
    \label{fig:ExtendedFigure_04} 
\end{figure*}

Finally, we assume the continuity model for the SFH. This is the only SFH that we assume for \texttt{Prospector}, since the two parametric SFHs (constant model and delayed-tau model) are unable to reproduce the observed photometry without invoking additional free parameters (e.g., the escape fraction, or the shape of the diffuse dust attenuation curve). The inclusion of additional free parameters, such as the escape fraction, reduces the inferred stellar masses by roughly $0.5\ \mathrm{dex}$ without affecting any of the other inferred physical parameters. This SFH has two free parameters corresponding to the total stellar mass formed ($\mathrm{log}_{10}[M_{\ast}/M_{\odot}]$, assuming a uniform prior with $\mathrm{min} = +6.0$ and $\mathrm{max} = +12.0$) and the stellar metallicity ($\mathrm{log}_{10}[Z_{\ast}/Z_{\odot}]$, assuming a uniform prior with $\mathrm{min} = -3.0$ and $\mathrm{max} = +0.0$). It also has four additional free parameters, corresponding to the logarithm of the ratio of the SFRs in the five adjacent time-bins ($R_{\mathrm{SFR}}$, assuming Student's $t$-distribution prior with $\mu = 0.0, \sigma = 0.3$). These time-bins are spaced at lookback times of $0-3$, $3-10$, $10-30$, $30-100$, and $100-t_{\mathrm{univ}}$ million years, which assumes the SFH starts at the formation redshift $z_{\mathrm{form}} = 20$. This physical model has eight free parameters, which should be compared to the eleven photometric detections that we have for \mbox{JADES-GS-z14-0}.

\begin{table}
    \centering
    \renewcommand{\arraystretch}{1.4}
    \begin{tabular}{l||c}
    \hline
    \hline
    & \multicolumn{1}{c}{Assumed Physical Model with \texttt{Prospector}} \\
    \hline
    Inferred Property & Continuity SFH \\
    \hline
    $\mathrm{log}_{10}\left( M_{\ast}/M_{\odot} \right)$ & ${9.38}_{-0.02}^{+0.02}$ \\
    $\mathrm{log}_{10}\left( Z_{\ast}/Z_{\odot} \right)$ & ${0.08}_{-0.02}^{+0.02}$ \\
    $t_{\ast}/\mathrm{Myr}$ & ${83.1}_{-3.0}^{+4.0}$ \\
    $\mathrm{SFR}_{10}/\left[ M_{\odot}/\mathrm{yr} \right]$ & ${0.02}_{-0.02}^{+0.08}$ \\
    $\Sigma_{\mathrm{SFR}_{10}}/\left[ M_{\odot}/\mathrm{yr}/\mathrm{pc}^{2} \right]$ & ${0.05}_{-0.05}^{+0.20}$ \\
    $\mathrm{sSFR}_{10}/\mathrm{Gyr}^{-1}$ & ${0.01}_{-0.01}^{+0.03}$ \\
    $A_{V}/\left[\mathrm{AB\ mag}\right]$ & ${0.014}_{-0.008}^{+0.008}$ \\
    $\mathrm{log}_{10}\left( U \right)$ & ${-2.1}_{-0.5}^{+0.4}$ \\
    $\mathrm{EW}_{\mathrm{[OIII]}+\mathrm{H}\beta}/\Angstrom$ & ${2}_{-17}^{+16}$ \\
    \hline
    \end{tabular}
    \caption{\textbf{Inferred physical properties of \mbox{JADES-GS-z14-0} from SED modeling with \texttt{Prospector}.} Columns: (1) name and units of the inferred property and (2) measurements using the continuity SFH model. Reported measurements are median values and $1\sigma$ confidence intervals.}
    \label{tab:Properties_Prospector}
\end{table}

Similar to Figure~\ref{fig:Figure_02}, Extended Data Figure~\ref{fig:ExtendedFigure_04} shows the SED modeling for JADES-GS-z14-0 with \texttt{Prospector}. In the upper panel, the fiducial model fitting photometry is used to constrain the various SED models. The median of these models is the gray line and unfilled squares, while the $1\sigma$ confidence interval is the shaded region. Table~\ref{tab:Properties_Prospector} reports the 16th, 50th, and 84th percentiles for the inferred physical properties. In the lower panel, the median model photometry is compared with the measured fluxes and uncertainties ($\chi$), while the total $\chi^{2}$ value is reported on the right. \texttt{Prospector} suggests that the excess flux in F770W relative to F444W is from stellar continuum rather than nebular emission line contributions, producing strong Balmer breaks in the models, which we find unlikely for the following reasons:
\begin{enumerate}
    \item[\textbf{(\#1.)}] \texttt{Prospector} predicts stellar masses ($M_{\ast} \approx 10^{9.4}\ M_{\odot}$) that are nearly an order of magnitude larger than the \texttt{BAGPIPES} models ($M_{\ast} \approx 10^{8.7}\ M_{\odot}$). \citetm{Robertson:2024} demonstrated that halos with virial masses of $M_{\ast} \approx 10^{9.8-9.9}\ M_{\odot}$ have comparable abundances to nine galaxy candidates at $z = 12-15$, including \mbox{JADES-GS-z14-0}. This implies that the \texttt{Prospector} predicted stellar mass is only three times smaller than the predicted halo mass through abundance matching, which is smaller than the predicted stellar-to-halo mass relation at $z \approx 14$ \citem{Behroozi:2020}.
    \item[\textbf{(\#2.)}] Theoretical predictions from the First Light And Reionization Epoch Simulations (FLARES)\citem{Wilkins:2023a} suggest that galaxies with stellar masses of $M_{\ast} \gtrsim 10^{9.0}\ M_{\odot}$ only exist at $z < 14$, while galaxies with $M_{\ast} \gtrsim 10^{9.5}\ M_{\odot}$ only exist at $z < 13$. Additionally, theoretical predictions from the IllustrisTNG and THESAN projects \citem{Kannan:2023} suggest that galaxies with $M_{\ast} \gtrsim 10^{9.0}\ M_{\odot}$ only exist at $z < 12$. Thus, cosmological galaxy formation simulations do not predict any galaxies at $z \approx 14$ with stellar masses that are comparable to those predicted by the \texttt{Prospector} models.
    \item[\textbf{(\#3.)}] The issues related to the extreme stellar masses predicted by \texttt{Prospector} are exacerbated by the extreme stellar age predictions ($t_{\ast} \approx 80-100$ million years, corresponding to formation redshifts of $z \approx 18-20$). Given that cosmological simulations are unable to produce galaxies at $z \approx 14$ with stellar masses that are comparable to those predicted by the \texttt{Prospector} models, they certainly cannot produce those same galaxies at $z \approx 18-20$. On the other hand, the stellar masses, stellar ages, specific star-formation rates, and diffuse dust attenuations predicted by the \texttt{BAGPIPES} models are consistent with theoretical predictions from FLARES \citem{Wilkins:2023a}.
\end{enumerate}

If they are correct, the physical properties predicted by the \mbox{\texttt{Prospector}} models would have radical implications for models of galaxy evolution in the early Universe, since cosmological simulations do not predict any galaxies at the redshift frontier with these extreme inferred stellar masses and stellar ages. For the reasons outlined above, we consider the solutions from \texttt{BAGPIPES} to be more plausible than the solutions from \texttt{Prospector}, but we are unable to reject either of these solutions on a formal basis. As a reminder, we attribute any differences in the predicted physical properties between \texttt{BAGPIPES} and \texttt{Prospector} to differences in the assumed stellar libraries, in addition to the stellar evolutionary tracks and isochrones.

\begin{figure*}
    \centering
    \includegraphics[width=0.9\textwidth]{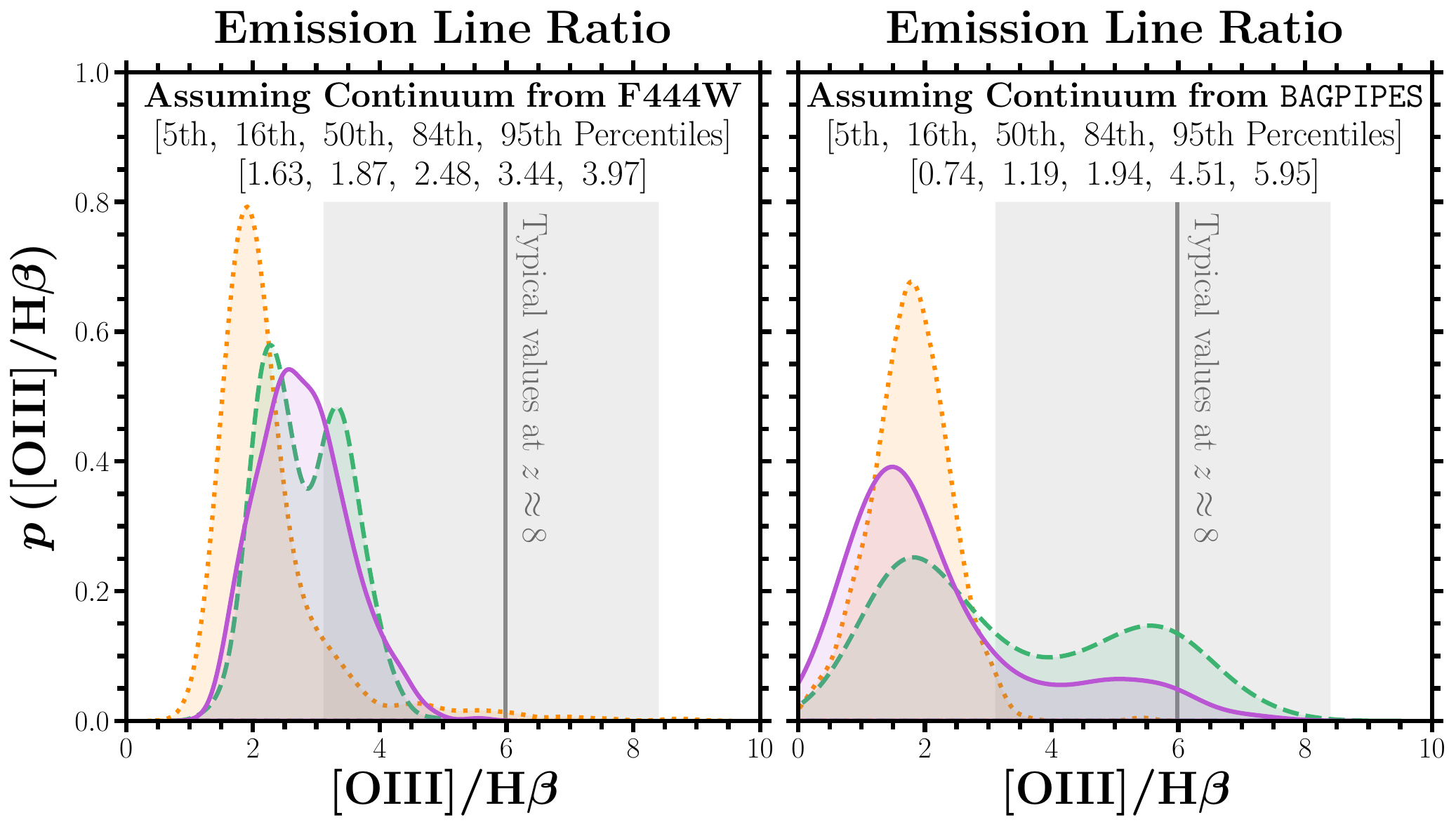}
    \caption{\textbf{\boldmath$\mathrm{[OIII]}/\mathrm{H}\beta$ emission line ratio predictions for JADES-GS-z14-0.} The physical properties inferred from SED modeling with \texttt{BAGPIPES} are used to predict the emission line ratio $\mathrm{[OIII]}/\mathrm{H}\beta$. This requires an assumption about the underlying continuum at rest-frame optical wavelengths. The left panel assumes the continuum is flat and consistent with the measured flux density in F444W, while the right panel assumes the posterior distribution of continuum levels from \texttt{BAGPIPES}.  Various SED models with different SFH assumptions (constant in purple, delayed-tau in green, and continuity in orange) are implemented. We report the 5th, 16th, 50th, 84th, and 95th percentiles after combining the posterior distributions from the various SED models. For comparison, we include measurements for a sample of galaxies at $z \approx 8$ with $\mathrm{H}\beta$ line flux measurements. The median of this sample is the gray lines, while the $1\sigma$ confidence interval is the shaded regions.}
    \label{fig:ExtendedFigure_05} 
\end{figure*}

Bayesian SED fitting requires numerous assumptions to interpret observations and infer physical properties for galaxies. We have explored the impact of adopting different SFHs and fitting codes, where the former had very little impact on the derived stellar population properties, while the latter had significant impact. Another assumption which can have significant impact on the derived stellar population properties is the IMF. In particular, it is likely that the formation of low-mass stars (i.e., those with stellar masses $M_{\ast} < 1-3\ M_{\odot}$) is strongly suppressed at high redshifts. This is caused by the high temperature of the cosmic microwave background \citem{Chon:2022} and the low metallicity of galaxies at $z > 10$ \citem{Steinhardt:2023, Chon:2024}. To explore the impact of these physical processes, we utilize the Bayesian SED fitting code \texttt{Prospector}, since \texttt{BAGPIPES} does not allow changes to the assumed IMF. With regard to top heavy IMFs,  \texttt{Prospector} provides the more demanding test because it fits the stellar continuum with older, low-mass stars whose formation is likely to be suppressed. For simplicity, we will assume the same Kroupa IMF \citem{Kroupa:2002} as before, but vary the lower and upper bounds of the mass range. Increasing the lower bound from $0.08\ M_{\odot}$ to $1\ M_{\odot}$ ($3\ M_{\odot}$) decreases the inferred stellar masses by roughly $0.3\ \mathrm{dex}$ ($0.4\ \mathrm{dex}$), while increasing the upper bound from $120\ M_{\odot}$ to $300\ M_{\odot}$ has no effect on the inferred stellar masses. These results suggest that the reported stellar masses based on a local IMF can be overestimated by up to a factor of three, without affecting the resulting SED. Recent work has found similar conclusions by changing assumptions about the IMF \citem{Woodrum:2024}.

\subsection{Nebular Emission Line Flux Predictions}
\label{Subsection:NebularEmissionLineFluxPredictions}

The aforementioned stellar population and dust properties provide predictions for the strength of the Balmer hydrogen recombination lines $\mathrm{H}\beta$ and $\mathrm{H}\alpha$. With an estimate of the strength of H$\beta$, it is possible to determine the remaining excess flux in F770W relative to F444W, which can be attributed to the metallic collisionally excited lines $\mathrm{[OIII]}\lambda\lambda4959,5007$. These predictions require a series of steps and assumptions  to convert the inferred star-formation rates into emission line flux predictions then broadband flux predictions.

The first step in this process involves converting the \texttt{BAGPIPES} derived star-formation rate into an $\mathrm{H}\beta$ line luminosity. To accomplish this, we proceed analogously to the fundamental relations derived by \citetm{Kennicutt:1998}. We compare \texttt{BAGPIPES} derived star-formation rates with observed $\mathrm{H}\beta$ line luminosities for a sample of $N = 27$ galaxies at $z \approx 8$ with $\mathrm{H}\beta$ line flux measurements from NIRSpec/PRISM. These measurements, along with Kron photometry convolved to the F444W PSF, are from the JADES Data Release 3 (DR3)\citem{D'Eugenio:2024}. We select all sources in GOODS-S with NIRCam coverage and signal-to-noise ratios $\mathrm{S/N} > 3$ for the $\mathrm{H}\beta$ line flux measurements. The measured photometry is used to constrain the various SED models with \texttt{BAGPIPES}, with procedures identical to those used for \mbox{JADES-GS-z14-0}, assuming the same three SFHs and concatenating the results from each. This produces a distribution of calibrations for each of the galaxies, which we can combine to get the full distribution of calibrations for the entire sample. They produce the following calibration:
\begin{equation}
    \label{Equation_1}
    C_{\mathrm{H}\beta} = \frac{L_{\mathrm{H}\beta}/[\mathrm{ergs/s}]}{\mathrm{SFR}_{10}/[M_{\odot}/\mathrm{yr}]} = 9.1 \pm 5.6 \times 10^{40}.
\end{equation}
The reported calibration and corresponding uncertainty come from taking the sigma-clipped mean and standard deviation of the full distribution of calibrations. The quoted uncertainty reflects the scatter around the average relation for individual galaxies. The dominant term in this uncertainty comes from the derived star-formation rates. Since the derived star-formation rate uncertainties are propagated throughout, we do not propagate the calibration uncertainty for the remaining calculations. The reported calibration results in a predicted $\mathrm{H}\beta$ line luminosity $L_{\mathrm{H}\beta} \approx 230_{-150}^{+150} \times 10^{40}\ \mathrm{ergs/s}$.

The second step in this process involves converting the $\mathrm{H}\beta$ line luminosity into a $\mathrm{H}\beta$ line flux. This requires a luminosity distance, $d_{L} = 4.803 \times 10^{29}\ \mathrm{cm}$ at $z = 14.32$, which is dependent on the assumed cosmology. This results in a predicted $\mathrm{H}\beta$ line flux $F_{\mathrm{H}\beta} \approx 7.9_{-1.8}^{+1.7} \times 10^{-19}\ \mathrm{ergs/s/cm}^{2}$.

The third step in this process involves converting the $\mathrm{H}\beta$ line flux into the equivalent signal in the F770W band. This requires the effective bandwidth of the F770W filter, $W_{\mathrm{\nu,\,\mathrm{F770W}}} = 1.00 \times 10^{13}\ \mathrm{Hz}$, as measured in frequency space and reported in the JWST User Documentation for the MIRI Filters and Dispersers. This results in a predicted $\mathrm{H}\beta$ F770W flux density $f_{\mathrm{H}\beta,\,\mathrm{F770W}} \approx 7.9_{-1.8}^{+1.7}\ \mathrm{nJy}$.

The final step in this process involves comparing the $\mathrm{H}\beta$ F770W flux density with the total nebular emission line contribution to F770W. These nebular contributions are equal to the F770W flux density minus the total continuum contribution to F770W. For the continuum contribution to F770W, we first assume the continuum is flat at rest-frame optical wavelengths and consistent with the measured flux density in F444W. We measure an excess flux of $\Delta f = 27.5 \pm 5.6\ \mathrm{nJy}$ in F770W relative to F444W, which results in a predicted line ratio $\mathrm{[OIII]}/\mathrm{H}\beta \approx 2.5_{-0.6}^{+0.9}$, as illustrated in the left panel of Extended Data Figure~\ref{fig:ExtendedFigure_05}. The quoted uncertainties on the line ratio include uncertainties on the star-formation rates, but not on the rest-frame optical continuum. To account for uncertainties in the rest-frame optical continuum, we further assume the posterior distribution of continuum levels from \texttt{BAGPIPES}. This results in a predicted line ratio $\mathrm{[OIII]}/\mathrm{H}\beta \approx 1.9_{-0.7}^{+2.6}$, as illustrated in the right panel of Extended Data Figure~\ref{fig:ExtendedFigure_05}. Regardless of assumption about the {rest-frame optical} continuum, we find that the excess flux at $7.7\ \mu\mathrm{m}$ includes a substantial contribution from the $\mathrm{[OIII]}\lambda\lambda4959,5007$ lines. This line ratio is smaller than typical values observed for galaxies at $z \approx 8$, as shown in Extended Data Figure~\ref{fig:ExtendedFigure_05} with individual galaxies that have $\mathrm{H}\beta$ line flux measurements from JADES DR3 \citem{D'Eugenio:2024}. 

The line ratio $\mathrm{[OIII]}/\mathrm{H}\beta$ is found to be correlated but degenerate with the gas-phase metallicity, with an additional, secondary dependence on the ionization and excitation states of the nebular gas. The turnover in the relation between $\mathrm{[OIII]}/\mathrm{H}\beta$ and gas-phase metallicity occurs roughly around $Z_{\ast} \approx 25\%\ Z_{\odot}$ \citem{Wilkins:2023b}, where lower values correspond to lower stellar masses, and higher values correspond to higher stellar masses. Given the inferred stellar masses and metallicites from the SED fitting for JADES-GS-z14-0 ($M_{\ast} \lesssim 10^{9}\ M_{\odot}$ and $Z_{\ast} \lesssim 10\%\ Z_{\odot}$), we assume the lower value of the gas-phase metallicity to break the double-valued degeneracy between $\mathrm{[OIII]}/\mathrm{H}\beta$. The inferred stellar masses and metallicities for the individual galaxies that have $\mathrm{H}\beta$ line flux measurements from JADES DR3 \citem{D'Eugenio:2024} are similar to those derived for JADES-GS-z14-0, so we can make the same assumption for those galaxies. Thus, given that the line ratio $\mathrm{[OIII]}/\mathrm{H}\beta$ for JADES-GS-z14-0 is smaller than typical values observed for galaxies at $z \approx 8$, these results suggest a smaller gas-phase metallicity for JADES-GS-z14-0 when compared to galaxies at $z \approx 8$.

\section*{Supplementary Information}

\textbf{Any correspondence and requests for materials should be addressed to Jakob M. Helton (\href{mailto:jakobhelton@arizona.edu}{jakobhelton@arizona.edu}).}

\subsection*{Data Availability}

The NIRCam data that support the findings of this study are publicly available at \href{https://archive.stsci.edu/hlsp/jades}{https://archive.stsci.edu/hlsp/jades}. The MIRI data that support the findings of this study will be made available in a future release; advanced access may be granted on reasonable request to the corresponding author.

%%%

\subsection*{Code Availability}

The \texttt{AstroPy} \citem{Astropy:2013, Astropy:2018} software suite is publicly available, as is \texttt{BAGPIPES} \citem{Carnall:2018}, \texttt{Cloudy} \citem{Byler:2017}, \texttt{dynesty} \citem{Speagle:2020}, \texttt{ForcePho} (\mbox{B. D. Johnson et al., in preparation}), \texttt{FSPS} \citem{Conroy:2009, Conroy:2010}, \texttt{GALFIT} \citem{Peng:2002, Peng:2010}, \texttt{Matplotlib} \citem{Matplotlib:2007}, \texttt{nautilus} \citem{Lange:2023}, \texttt{NumPy} \citem{NumPy:2011, NumPy:2020}, \texttt{Pandas} \citem{Pandas:2022}, \mbox{\texttt{photutils}} \citem{Bradley:2022}, \texttt{Prospector} \citem{Johnson:2021}, \texttt{python-FSPS} \citem{Foreman-Mackey:2014}, \texttt{SciPy} \citem{SciPy:2020}, \texttt{seaborn} \citem{Waskom:2021}, \texttt{TinyTim} \citem{Krist:2011}, and \texttt{WebbPSF} \citem{Perrin:2014}.

%%%

\subsection*{Acknowledgments}

J.M.H., G.H.R., S.A., Z.H., D.J.E., P.A.C., E.E., B.D.J., M.J.R., B.R., F.S., and C.N.A.W. are supported by NASA contracts NAS5-02105 and NNX13AD82G to the University of Arizona. D.J.E. is supported as a Simons Investigator. S.C. acknowledges support by European Union’s HE ERC Starting Grant No. 101040227 - WINGS. W.M.B. and J.S. acknowledges support by the Science and Technology Facilities Council (STFC), ERC Advanced Grant 695671 ``QUENCH''. A.J.B., J.C., G.C.J., and A.S. acknowledge funding from the ``FirstGalaxies'' Advanced Grant from the European Research Council (ERC) under the European Union’s Horizon 2020 research and innovation programme (Grant agreement No. 789056). F.D.E., R.M., and J.W. acknowledges support by the Science and Technology Facilities Council (STFC), by the ERC through Advanced Grant 695671 ``QUENCH'', and by the UKRI Frontier Research grant RISEandFALL. R.M. also acknowledges funding from a research professorship from the Royal Society. P.G.P.-G. acknowledges support from grant PID2022-139567NB-I00 funded by Spanish Ministerio de Ciencia e Innovaci\'on MCIN/AEI/10.13039/501100011033, FEDER, UE. S.T. acknowledges support by the Royal Society Research Grant G125142. L.W. acknowledges support from the National Science Foundation Graduate Research Fellowship under Grant No. DGE-2137419. The research of C.C.W. is supported by NOIRLab, which is managed by the Association of Universities for Research in Astronomy (AURA) under a cooperative agreement with the National Science Foundation.

This work made use of the {\it lux} supercomputer at UC Santa Cruz which is funded by NSF MRI grant AST 1828315, as well as the High Performance Computing (HPC) resources at the University of Arizona which is funded by the Office of Research Discovery and Innovation (ORDI), Chief Information Officer (CIO), and University Information Technology Services (UITS). We respectfully acknowledge the University of Arizona is on the land and territories of Indigenous peoples. Today, Arizona is home to 22 federally recognized tribes, with Tucson being home to the O’odham and the Yaqui. Committed to diversity and inclusion, the University strives to build sustainable relationships with sovereign Native Nations and Indigenous communities through education offerings, partnerships, and community service.

%%%

\subsection*{Author Contributions Statement}

J.M.H. and G.H.R. led the writing of this paper. J.M.H., G.H.R., S.A., Z.W., D.J.E., Z.J., J.L., P.G.P.-G., and I.S. contributed to the MIRI imaging reduction and the modeling of the MIRI photometry. D.J.E., P.A.C., B.D.J., M.J.R., B.R., S.T., and C.N.A.W. contributed to the NIRCam imaging reduction and the modeling of the NIRCam photometry. S.Carniani, A.J.B., S.Charlot, J.C., F.D.E., R.M., and J.W. contributed to the NIRSpec data reduction and the modeling of the NIRSpec spectroscopy. J.M.H., G.H.R., S.A., K.N.H., S.Carniani., Z.J., S.Charlot, J.C., F.D.E., B.D.J., G.C.J., J.L., B.R., J.S., A.S., S.T., L.W., and J.W. contributed to the modeling and interpretation of the spectral energy distribution. K.N.H. contributed to the photometric redshift determination. F.S. contributed to the lensing magnification estimation. B.R. contributed to the imaging data visualization. K.N.H., M.J.R., B.R., C.C.W., and C.N.A.W contributed to the pre-flight NIRCam imaging data challenges. G.H.R., S.A., D.J.E., A.J.B., R.M., M.J.R., B.R., C.N.A.W., and C.W. contributed to the design of the MIRI, NIRCam, and NIRSpec observations. G.H.R., S.A., K.N.H., R.B., E.E., M.J.R., F.S., C.C.W., and C.N.A.W. contributed to the design, construction, and commissioning of MIRI, NIRCam, and NIRSpec. W.M.B. and Y.Z. contributed comments to the paper.

%%%

\subsection*{Competing Interests Statement}

The authors declare no competing interests.

%%%%%%%%%%%%%%%%%%%% REFERENCES %%%%%%%%%%%%%%%%%%

\bibliographym{manuscript}

\section*{Affiliations}
\noindent
\hypertarget{inst:Steward}$^{1}$Steward Observatory, University of Arizona, 933 N. Cherry Ave., Tucson, AZ 85721, USA
\\
\hypertarget{inst:CfA}$^{2}$Center for Astrophysics $|$ Harvard \& Smithsonian, 60 Garden St., Cambridge, MA 02138, USA
\\
\hypertarget{inst:SNS}$^{3}$Scuola Normale Superiore, Piazza dei Cavalieri 7, I-56126 Pisa, Italy
\\
\hypertarget{inst:Kavli}$^{4}$Kavli Institute for Cosmology, University of Cambridge, Madingley Road, Cambridge CB3 0HA, UK
\\
\hypertarget{inst:Cav}$^{5}$Cavendish Laboratory, University of Cambridge, 19 JJ Thomson Avenue, Cambridge CB3 0HE, UK
\\
\hypertarget{inst:ESAC}$^{6}$European Space Agency (ESA), European Space Astronomy Centre (ESAC), Camino Bajo del Castillo s/n, 28692 Villanueva de la Ca\~{n}ada, Madrid, Spain
\\
\hypertarget{inst:Oxford}$^{7}$Department of Physics, University of Oxford, Denys Wilkinson Building, Keble Road, Oxford OX1 3RH, UK
\\
\hypertarget{inst:IAP}$^{8}$Sorbonne Universit\'{e}, CNRS, UMR 7095, Institut d'Astrophysique de Paris, 98 bis bd Arago, 75014 Paris, France
\\
\hypertarget{inst:UCL}$^{9}$Department of Physics and Astronomy, University College London, Gower Street, London WC1E 6BT, UK
\\
\hypertarget{inst:CAB}$^{10}$Centro de Astrobiolog\'{i}a (CAB), CSIC–INTA, Cra. de Ajalvir Km. 4, 28850- Torrej\'{o}n de Ardoz, Madrid, Spain
\\
\hypertarget{inst:UCSC}$^{11}$Department of Astronomy and Astrophysics University of California, Santa Cruz, 1156 High St., Santa Cruz, CA 96054, USA
\\
\hypertarget{inst:NOIRLab}$^{12}$NSF's National Optical-Infrared Astronomy Research Laboratory, 950 N. Cherry Ave., Tucson, AZ 85719, USA
\\
\hypertarget{inst:NRC}$^{13}$NRC Herzberg, 5071 W. Saanich Rd., Victoria, BC V9E 2E7, Canada

% Don't change these lines
% \bsp% typesetting comment

\end{document}